\documentclass[nofootinbib]{revtex4-2}

\usepackage{scrextend}
\usepackage{graphicx}
\usepackage{indentfirst}
\usepackage{latexsym} 
\usepackage{multirow}
\usepackage{epsfig}
\usepackage{hyperref}
\usepackage{color}
\usepackage[usenames,dvipsnames]{xcolor}
\usepackage{amssymb}
\usepackage{amsfonts}
\usepackage{amsmath}
\usepackage{bm}
\usepackage{mathrsfs}
\usepackage{cleveref}
\usepackage{algorithm}
\usepackage{algorithmic}
\usepackage{subcaption}
\algsetup{indent=2em}

\usepackage[normalem]{ulem}
\useunder{\uline}{\ul}{}

\newcounter{comment}

\makeatletter
\newcommand{\multiline}[1]{%
  \begin{tabularx}{\dimexpr\linewidth-\ALG@thistlm}[t]{@{}X@{}}
    #1
  \end{tabularx}
}
\newcommand{\bea}{\begin{eqnarray}}
\newcommand{\eea}{\end{eqnarray}}
\makeatother

\newcommand{\be}{\begin{equation}}
\newcommand{\ee}{\end{equation}}

\def\addfamafconicet{CONICET, Córdoba, Argentina\\
Facultad de Matem\'atica, Astronom\'ia, F\'isica y Computaci\'on,\\
 Universidad Nacional de C\'ordoba, (5000) C\'ordoba, Argentina}

\def\addIsistan{ISISTAN-CONICET Research Institute, \\UNICEN University, (7000) Tandil, Buenos Aires, Argentina}

\def\addFacet{Facultad de Ciencias Exactas y Tecnolog\'ia, Universidad Nacional de Tucum\'an }

\begin{document}

\title{Hyperparameter optimization of hp-greedy reduced basis for gravitational wave surrogates}

\author{Franco Cerino\textsuperscript{1}, J. Andrés Diaz-Pace\textsuperscript{2}, Emmanuel A. Tassone\textsuperscript{1}, Manuel Tiglio\textsuperscript{1}, Atuel Villegas\textsuperscript{3}}
\affiliation{\textsuperscript{1}\addfamafconicet}
\affiliation{\textsuperscript{2}\addIsistan}
\affiliation{\textsuperscript{3}\addFacet}

\begin{abstract}
In a previous work~\cite{Cerino:2022dhr} we introduced, in the context of gravitational wave science, an initial study on an automated domain-decomposition approach for reduced basis through hp-greedy refinement. The approach constructs local reduced bases of lower dimensionality than global ones, with the same or higher accuracy. These ``light'' local bases should imply both faster evaluations when predicting new waveforms and faster data analysis, in particular faster statistical inference (the forward and inverse problems, respectively). 
In this approach, however, we have previously found important dependence on several hyperparameters, which do not appear in global reduced basis. This naturally leads to the problem of hyperparameter optimization (HPO), which is the subject of this paper. 
We tackle the problem through a Bayesian optimization, and show its superiority when compared to grid or random searches. We find that for gravitational waves from the collision of two spinning but non-precessing black holes, for the same accuracy, local hp-greedy reduced bases with HPO have a lower dimensionality of up to $4 \times$ for the cases here studied, depending on the desired accuracy. This factor should directly translate in a parameter estimation speedup, for instance. Such acceleration might help in the near real-time requirements for electromagnetic counterparts of gravitational waves from compact binary coalescences. In addition, we find that the Bayesian approach used in this paper for HPO is two orders of magnitude faster than, for example, a grid search, with about a $100 \times$ acceleration.  
The code developed for this project is available as open source from public repositories. \footnote{This paper is an invited contribution to the Special Issue ``Recent Advances in Gravity: A Themed Issue in Honor of Prof. Jorge Pullin on his 60th Anniversary''.}
\end{abstract}

\maketitle

\section{Introduction} \label{sec:intro}
For several problems, in particular data-driven ones as is the case of this paper, surrogate models have proved useful to make both predictions and analyses on new data computationally faster. These models are constructed by learning from a limited dataset, obtained for example from high-fidelity simulations or experiments. This paper uses the reduced basis approach to construct surrogates, for a review see Ref.~\cite{tiglio2022reduced}.


Parameter estimation (PE) of the source of gravitational waves is a key aspect of gravitational wave (GW) science~\cite{mandic2012parameter,isi2019hierarchical,lange2018rapid,lynch2017information,mandel2014parameter,mandel2019extracting,usman2019constraining,van2009parameter}; its goal is to infer the properties of, for example, the black holes or neutron stars involved in a binary collision~\cite{fishbach2020does,cornish2021rapid,berry2015parameter,biscoveanu2020quantifying,bizouard2021inference,banagiri2020constraining,coughlin2019multimessenger,wysocki2019reconstructing}. Along this line, speeding up PE can enable the possibility of measuring  electromagnetic counterparts of gravitational waves in the presence of a neutron star~\cite{christensen2022parameter,jaranowski2012gravitational}. This counterpart refers to the electromagnetic signal(s) received after a gravitational wave.  Bayesian inference is a standard approach in PE~\cite{smith2021bayesian,breschi2021bayesian,chua2020learning,meyer2022computational,edwards2014bayesian,dupuis2005bayesian,talbot2019parallelized} and several software tools have been developed in GW science, such as LALInference~\cite{veitch2015parameter}, PyCBC~\cite{biwer2019pycbc} and Bilby~\cite{ashton2019bilby,romero2020bayesian}.

The main factors contributing to the PE computational costs are waveform evaluations and likelihood computations. One way to overcome the first one is through surrogate models. Analogously, likelihood evaluations can be sped up through the use of reduced order quadratures (ROQ)~\cite{antil2012two,Canizares:2013ywa,canizares2015accelerated},  which are based on reduced order models and the Empirical Interpolation method~\cite{barrault2004empirical}. Several efficiency improvements for PE have also been reported using standard Machine learning (ML) techniques~\cite{dax2021real,gabbard2019bayesian,green2020complete,green2020gravitational,george2017deep,alvares2021exploring,shen2021statistically}. Even though the acceleration of likelihood evaluations --and PE thereof-- using ROQ is significant, they are not yet enough to allow for the follow-up of electromagnetic counterparts. One further acceleration being proposed is the use of focused reduced order quadratures (FROQ)~\cite{Morisaki_2020}, which are built from a reduced basis in a neighborhood of the parameter region found by the trigger (detection) pipeline, as opposed to a global basis covering the whole parameter domain of physical possibilities. Since the parameter region is smaller, the basis has a lower dimensionality, and the cost of evaluating ROQs is linear with respect to the dimensionality of the basis. 
 
In a recent paper~\cite{Cerino:2022dhr} we proposed a divide-and-conquer approach to build accurate local reduced bases of low dimensionality in an automated  way, which can be seen as complementary to FROQ. More precisely, we use a data-driven version of hp-greedy reduced basis~\cite{EftangThesis} \footnote{The hp-greedy approach, as well as reduced basis, was originally introduced in the context of parameterized partial differential equations.}; a methodology that adaptively partitions the parameter space and builds local reduced bases. 
In that reference we emphasized that the hp-greedy approach has significant online speed improvements, given that it obtains a set of reduced bases with lower dimensionality and equal or higher accuracy than a global basis. At the same time, 
the hp-greedy approach is more complex than the standard reduced basis one. In particular, in~\cite{Cerino:2022dhr} we also found that there are  hyperparameters to which the resulting models are very sensitive and which do not appear (or are irrelevant) in the standard reduced basis framework. We have identified the two most relevant ones to optimize for: 
\begin{enumerate}
\item The seed $\hat{\Lambda}_0$ to initialize the greedy-reduced basis construction. This was largely unexpected, since it has been consistently reported in the literature in the past that it has no relevance in global reduced basis \footnote{Which can be intuitively understood in that case being the greedy approach a {\em global} optimization algorithm.}, see for example Figure 1 and its associated discussion in Ref.~\cite{Caudill:2011kv}.  
\item The maximum depth $l_{max}$ of the resulting tree (briefly described in the next section), which limits the number of recursive partitions. As with any tree in ML, deeper trees lead to higher accuracies when training but at the same time they risk overfitting. 
\end{enumerate}

The previous discussion motivates the subject of this paper: our approach of hp-greedy reduced basis requires an efficient search for an optimal choice of the hp-greedy hyperparameters. This is referred to as hyperparameter optimization (HPO) in the ML field. Here we follow a Bayesian approach; more precisely, Sequential Model-Based Optimization (SMBO) through a Tree-Structured Parzen Estimator (TPE)~\cite{NIPS2011_86e8f7ab}, as implemented in the OPTUNA open source package~\cite{optuna_2019}. 

The rest of the paper is organized as follows. In section \ref{sec:hp} we briefly review the hp-greedy reduced basis approach. In Section \ref{sec:hpo} we state the hyperparameter optimization problem and describe how we approach it using Bayesian optimization, SMBO and TPE. We also include a benchmark comparison of HPO with a function commonly used in optimization tests. In Section \ref{sec:results} we present our results for the collision of  two non-precessing aligned-spins black holes, the same physical setup that we used in our previous work~\cite{Cerino:2022dhr}. We close in Section \ref{sec:com} presenting the conclusions of this work and potential future paths. 
 
\section{hp-greedy reduced basis} \label{sec:hp}

The reduced basis-greedy approach finds a quasi-optimal --in a rigorous mathematical sense-- low-dimensional approximation of a dataset with respect to the Kolmogorov n-width~\cite{Binev10convergencerates,Devore2012}. When applied to gravitational waves, the dataset consists of waveforms parameterized, for example, by the mass and spin of each object in the case of a compact binary coalescence.
To further accelerate online evaluations and data analysis, a divide-and-conquer strategy can be pursued, which recursively partitions the parameter space and builds local reduced bases of lower dimensionality.  We therefore proposed~\cite{Cerino:2022dhr} a data-driven version of the {\it hp-greedy} approach~\cite{EftangThesis} as a way of improving the construction of reduced bases within gravitational wave science.  

As a summary, hp-greedy generalizes the standard reduced-basis framework, allowing for partitioning of the parameter space by combining reduced basis functions (p-refinement) with an adaptive grid  (h-refinement). Each subspace can be assigned to a node in a binary tree structure. If a subspace is partitioned, it is associated with a node of the tree and, after partitioning it, each of the obtained subspaces is associated with a children node. In ths way, the root of the tree represents the full parameter space. For more details see~\cite{Cerino:2022dhr}.  

\begin{figure}[h!]
\centering
\includegraphics[width=.8\columnwidth]{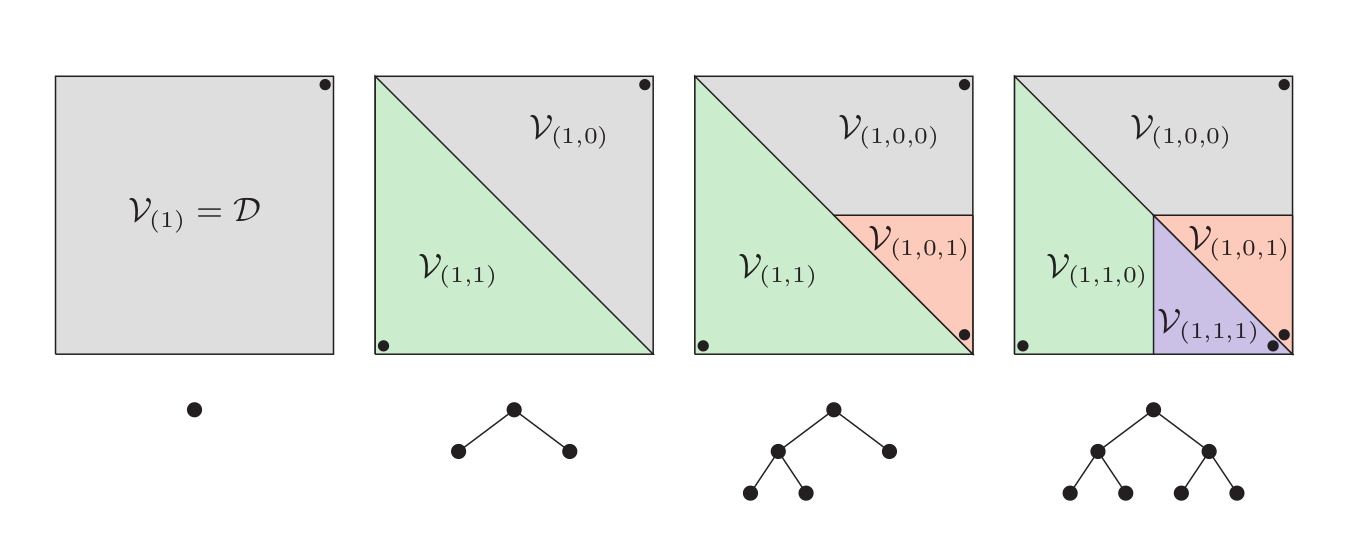}
\caption{Schematic domain decomposition and its associated tree representation. Figure from Ref.~\cite{EftangThesis}.}
\label{fig:hpgreedy}
\end{figure}

The process builds a set of local reduced bases with several stopping criteria for partitioning the parameter domain: a maximum dimension $n_{max}$ for each local basis, a maximum depth $l_{max}$ of the tree, and an accuracy threshold $\epsilon_{tol}$ as defined by Eq.~\ref{eq:error}. Until any of these stopping criteria is met, the domain gets partitioned, and reduced bases for the resulting subdomains are built. Figure~\ref{fig:hpgreedy} shows an example of a partition structure for a domain with $l_{max}=2$.

The hp-greedy approach is driven by the idea that if the greedy error decreases slowly, leading to a large number of iterations, domain partitioning can help to improve accuracy; which is similar in spirit to spectral elements in numerical solutions of partial differential equations~\cite{karniadakis2005spectralhp}. The choice of the partitioning is influenced by the rate of error reduction, which varies depending  on the problem. 
Numerical experiments have demonstrated the effectiveness of hp-greedy for gravitational waves, reducing basis dimensionality while maintaining accuracy (cf. Fig.~12 of Ref.~\cite{Cerino:2022dhr}). 

The algorithm involves performing a binary domain-decomposition, and for traversing the resulting tree one can take advantage of its structure; we discuss this in Section~\ref{sec:com}. hp-greedy is particularly useful for problems with physical discontinuities in the parameter space (cf. Sec. III of Reference~\cite{Cerino:2022dhr}).

An interesting finding of our experiments with hp-greedy is that the initial seed of the algorithm {\em does} affect the partitioning and subsequent reduced bases, and {\em significantly} impacts their {\em accuracy}. This differs from the standard global reduced basis approach, in which the seed choice is irrelevant, see for example Figure~1 of~\cite{Caudill:2011kv} and the corresponding discussion. Hence, the optimization of hyperparameters such as $l_{max}$, and the seed $\hat{\Lambda}_0$ turns out to be crucial in hp-greedy.  This observation is the key motivation for this paper.  The optimization task can be carried out through hyperparameter optimization in the ML sense. 

\section{Hyperparameter optimization} \label{sec:hpo}
An HPO problem can be stated as follows: given a cost function $f: X \rightarrow \mathbb{R}$ which returns the maximum validation error of a model trained with a combination of hyperparameters $\textbf{x} \in X$, the goal is to find $\textbf{x}^*$ such that

\[
\textbf{x}^* = arg \min_{\textbf{x} \in X} f(\textbf{x}) \, . 
\]
In our context, we are interested in the optimal combination of hyperparameters from a domain $X$ that gets the minimum representation error for a validation set. For a discussion about our results on test sets, see Section \ref{sec:performance}. 

In the hp-greedy approach, each value of the tuple $\textbf{x}$ represents a configuration of the two relevant hyperparameters for our scenario:
\[
\textbf{x} = (l_{max}, \hat{\Lambda}_0)\, , 
\]
for fixed values of $n_{max}$. We decided to keep the latter fixed since the evaluation cost of both a surrogate based on reduced bases and the computation of likelihoods using ROQ is linear with the dimensionality of the basis, so we place it in a different hierarchy. 

In practice, the cost function (which we have not defined yet) does not have a closed form expression, but instead it results from training a model and evaluating the representation error of a validation set. This aspect usually makes the optimization problem computationally expensive. 

Several HPO approaches have been and are being developed, and one of the driving motivations nowadays are deep neural networks. Two  well-known HPO techniques are grid and random searches, although they are often inefficient in computational terms since the whole space is blindly explored. A more promising technique in this regard is Bayesian optimization~\cite{7352306, https://doi.org/10.48550/arxiv.1012.2599}, which was chosen for our problem. It attempts to minimize the number of evaluations of $f$ to find $\textbf{x}^*$ and falls within the category of {\it Sequential Model-Based Optimization} (SMBO)~\cite{dewancker2015bayesian,NIPS2011_86e8f7ab}. In this work, we rely on  the \textit{Tree-Structured Parzen Estimator} (TPE)~\cite{NIPS2011_86e8f7ab} algorithm for Bayesian optimization, because it is one of the simplest algorithms. It works well with discrete search spaces, scales linearly with the number of dimensions and is optimized --as opposed to other methods such as Gaussian Processes (see \cite{NIPS2011_86e8f7ab} for more details). For the SMBO implementation, we used the Python package OPTUNA~\cite{optuna_2019}. 

In Section~\ref{sec:benchmark} we compare our results using Bayesian optimization with those of grid and random searches to quantify the advantages and computational savings of the former.

\subsection{Bayesian optimization} \label{sec:bay}
\begin{figure}[h!]
\centering
\includegraphics[width=.5\columnwidth]{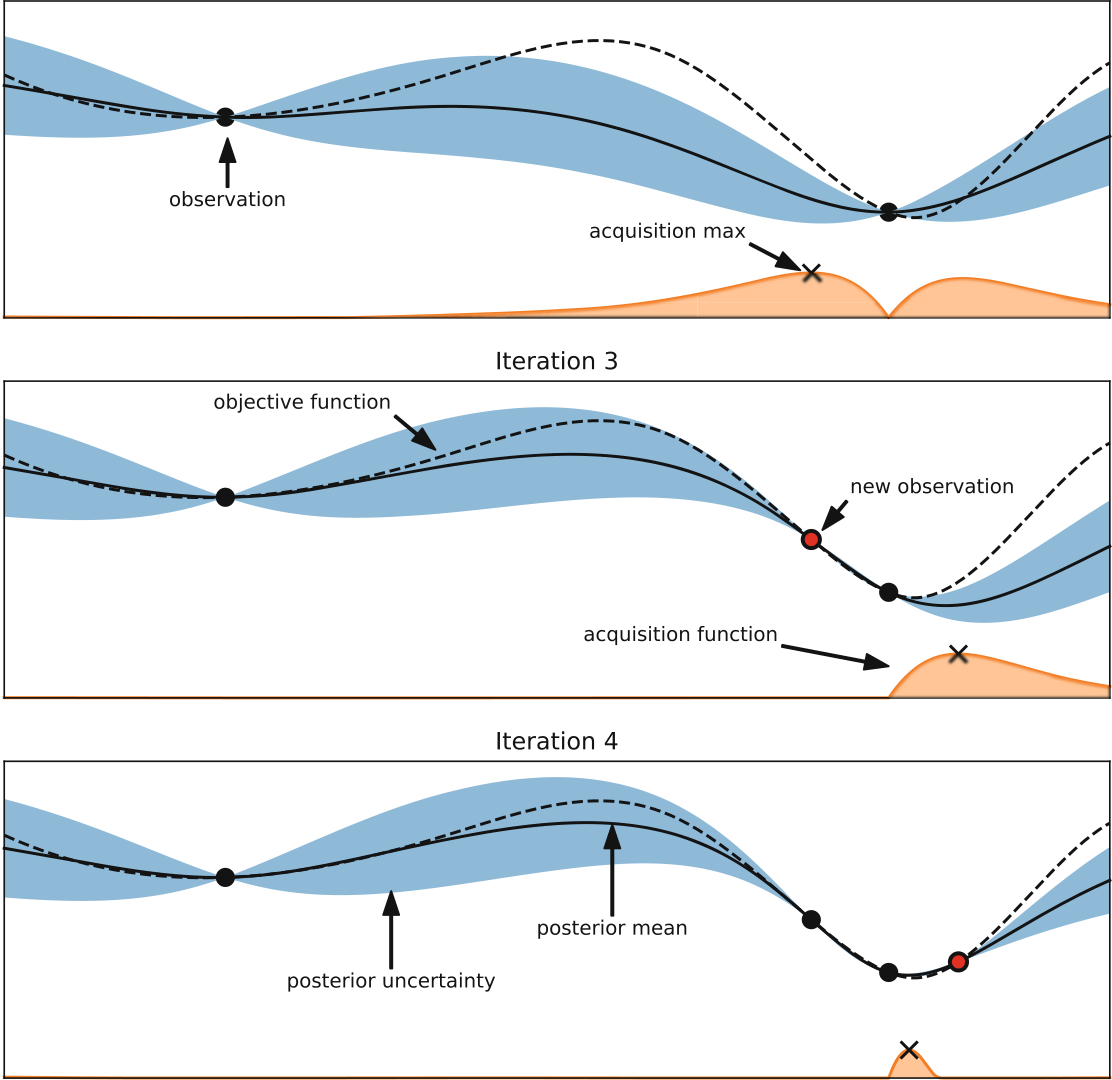}
\caption{Three iterations of a Bayesian optimization for a cost function with one parameter. The dashed line shows the actual cost function, and the solid one the mean value of a statistical model (in this case using Gaussian processes). The blue area shows the uncertainty of the model, which approaches zero at the points where the observations are made. Underneath, in orange, the acquisition function, which shows the next point to evaluate. Figure taken from~\cite{Feurer2019}.}
\label{fig:bayesian}
\end{figure}

Basically, Bayesian optimization is an adaptive method that uses the information from previous evaluations of the cost function $f$ to decide which value of $\textbf{x}$ should be used next, with the goal of reducing the number of necessary evaluations of $f$ to find a (in general, local) minimum, see Figure~\ref{fig:bayesian}. To explain the intuition behind this method, we begin with a description of SMBO. 

\subsection{Sequential Model-Based Optimization (SMBO)} \label{sec:smbo}

The general idea is to approximate the cost function $f$ with a substitute model $\mathcal{M}$. Let us start with a set of observations 
\be
D = \{(\textbf{x}^{(1)},y^{(1)}), \cdots, (\textbf{x}^{(k)},y^{(k)}) \} \, ,  \label{eq:D}
\ee
with $y^{(j)} = f(\textbf{x}^{(j)})$. Departing from this set, the substitute model $\mathcal{M}$ is adjusted. 
Next, using the predictions of the model, an acquisition function $S$ is maximized. This function chooses the next set of hyperparameters 
$\textbf{x}_i \in X$ to evaluate $f$, and the pair $(\textbf{x}_i, f(\textbf{x}_i))$ is added to the observation set $D$. After that, $\mathcal{M}$ is adjusted again, and the process is repeated for a fixed number of iterations. This procedure is captured by the pseudocode given in Algoritm \ref{alg:SMBO}. \

\begin{algorithm}[H]
\caption{\texttt{SMBO}}
\label{alg:SMBO}
\begin{algorithmic}[1]
\REQUIRE $f, X, S,\mathcal{M}$

\vspace{2mm}
$ f : \text{Objective function}$, \\
$ X: \text{Search space}$, \\
$\mathcal{M} : \text{Model,}$\\
$S: \text{Acquisition function.}$\\

\vspace{3mm}

\STATE $D =$ InitialSample$(f, X)$
\vspace{1mm}
\FOR{$i = 1, 2, ...$}
	\STATE $\mathcal{M} =$ AdjustModel$(D)$
	\STATE $\textbf{x}_{i} = arg \max_{\textbf{x}\in X} \mathcal{S}(\textbf{x}, \mathcal{M})$ .
	\STATE $y_i = f(\textbf{x}_i)$	
	\STATE $D = D \cup \{(\textbf{x}_i, y_i)\}$
\ENDFOR
\vspace{3mm}

\end{algorithmic}
\end{algorithm}

Using Bayes' theorem, if $P(y|\textbf{x})$ is the posterior probability, $P(y)$ the prior and $P(\textbf{x}|y)$ the likelihood, then 
 \[
 P(y|\textbf{x}) = \frac{P(\textbf{x}|y) \ P(y)}{P(\textbf{x})} \, . 
 \]
In a Bayesian approach to SMBO, $P(y|\textbf{x})$ is the prediction of the model, with $y$ being an evaluation of $f(\textbf{x})$. 

We mentioned that, for selecting the points to evaluate, an acquisition function $S$ is maximized. Several proposals exist for choosing the acquisition function. In this work, we use the \textit{Expected Improvement} (EI)~\cite{EI1} criterion: if $y^*$ is a reference value, then EI with respect to $y^*$ is defined as 
\begin{equation}
\label{eq:ei_def}
EI_{y^*}(\textbf{x}) := \int_{-\infty}^{\infty} \max(y^*-y,0) P(y|\textbf{x}) \ dy \, .
\end{equation}

\subsection{Tree-Structured Parzen Estimator} 
\label{sec:TPE}

The \textit{Tree-Structured Parzen Estimator} (TPE) ~\cite{NIPS2011_86e8f7ab} is a strategy to model $P(x_i|y)$ for each $x_i \in X_i$ (that is, each $x_i$ represents a different hyperparameter) from two distributions built using the observations $D$ \eqref{eq:D}:
\begin{equation}
\label{eq:tpe}
P(x_i|y) =
	\begin{cases}
		\ell (x_i) & \text{if } y <y^{*} \\
		g(x_i) & \text{if } y \geq y^{*} \, . 
	\end{cases}
\end{equation}
Here the densities $\{ \ell(x_i), g(x_i)\}$ are built from two sets $\{ D_{\ell}, D_g  \} \subset D$, such that $D_{\ell}$ has all the observations with $y < y^*$, $D_g$ the remaining ones, and $D = D_{\ell} \cup D_g$. 

The reference value $y^{*}$ is a quantile $\gamma \in(0,1)$ of the observed values, so that $P(y<y^{*}) = \gamma$. This means that $y^{*}$ is certain value between the best $y$  and worst $y$ found at some iteration (e.g. if $\gamma$ is equal to $0.5$, then $y^*$ is equal to the median of the observed values of $y$). 

Building $\ell(x_i)$ and $g(x_i)$ implies adjusting the model (line 3 in Algorithm~\ref{alg:SMBO}), and then using~\eqref{eq:tpe} in the definition of the expected improvement~\eqref{eq:ei_def}. In order to maximize the expected improvement (step 4 in Algorithm~\ref{alg:SMBO}) one has to choose a value of $x_i$ that maximizes the ratio $\ell(x_i)/g(x_i)$~\cite{NIPS2011_86e8f7ab}, 
\begin{equation}
x_i^* =  arg\max_{x_i} \left( \ell(x_i)/g(x_i) \right) \, .  
\end{equation}

In summary, the TPE algorithm constructs two probability density functions: i) $\ell(x_i)$ using ``good'' observations ($y < y^*$), and ii) $g(x_i)$ using ``bad'' observations ($y \geq y^*$). These functions are updated every time the objective function is evaluated (at every iteration of the algorithm), and the new $x_i$ is chosen by maximizing  $\ell(x_i)/g(x_i)$, implying that the new $x_i$ is much more likely to represent a good observation rather than a bad one. 
All density functions are constructed using Parzen estimators~\cite{parzen1962}. For each point $x_i$ a truncated normal distribution centered at that point is added. A way to think of a Parzen window is to visualize it as a smoothed histogram.
The choice of truncated normal distributions for the kernel function is done in the original TPE paper~\cite{NIPS2011_86e8f7ab}. For more details about the implementation of the Parzen estimator in OPTUNA see Ref.~\cite{10.1613/jair.1.13188}.

\subsection{A comparison between HPO, grid and random searches}
\label{sec:benchmark}

\begin{figure}[h!]
\centering
\includegraphics[width=.8\columnwidth]{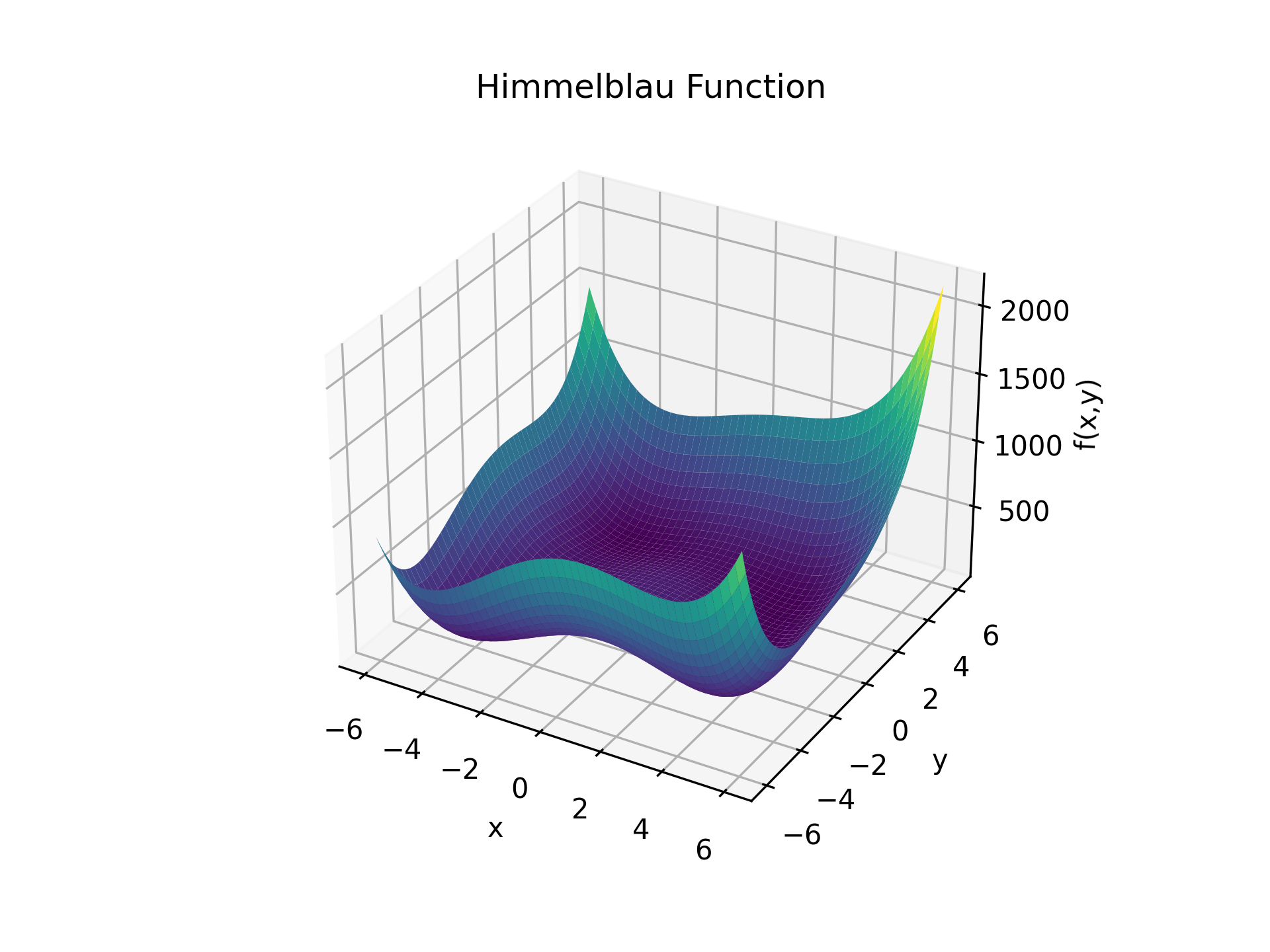}
\caption{Plot of the Himmelblau function. It has 4 global minima at $(-3.78,-3.28)$, $(-2.81,3.13)$, $(3.58,-1.85)$ and $(3,2)$, with the same value:  $f(x,y)=0$.}
\label{fig:grid_optimization_himmelblau}
\end{figure}

We compare grid and random searches against Bayesian optimization. To this end, let us consider the Himmelblau function, which is a widely used benchmark in optimization and ML, as the objective function to be minimized. This function is shown in Figure~\ref{fig:grid_optimization_himmelblau} and is defined as follows:
$$f(x,y) = (x^{2}+y-11)^{2} + (x+y^{2}-7)^{2}.$$

We assume that the Himmelblau function represents the accuracy of some ML model, where each input of the function represents a set of hyperparameters of the algorithm and the outcome is the resulting accuracy. The objective is to obtain a set of values that minimizes the function. We show results for three methods: grid and random searches, and Bayesian optimization. For each case, $100$ different evaluations were used. Then, we assess which one performed better. 

\begin{figure}[!htb]
\minipage{0.32\textwidth}
\includegraphics[width=\linewidth]{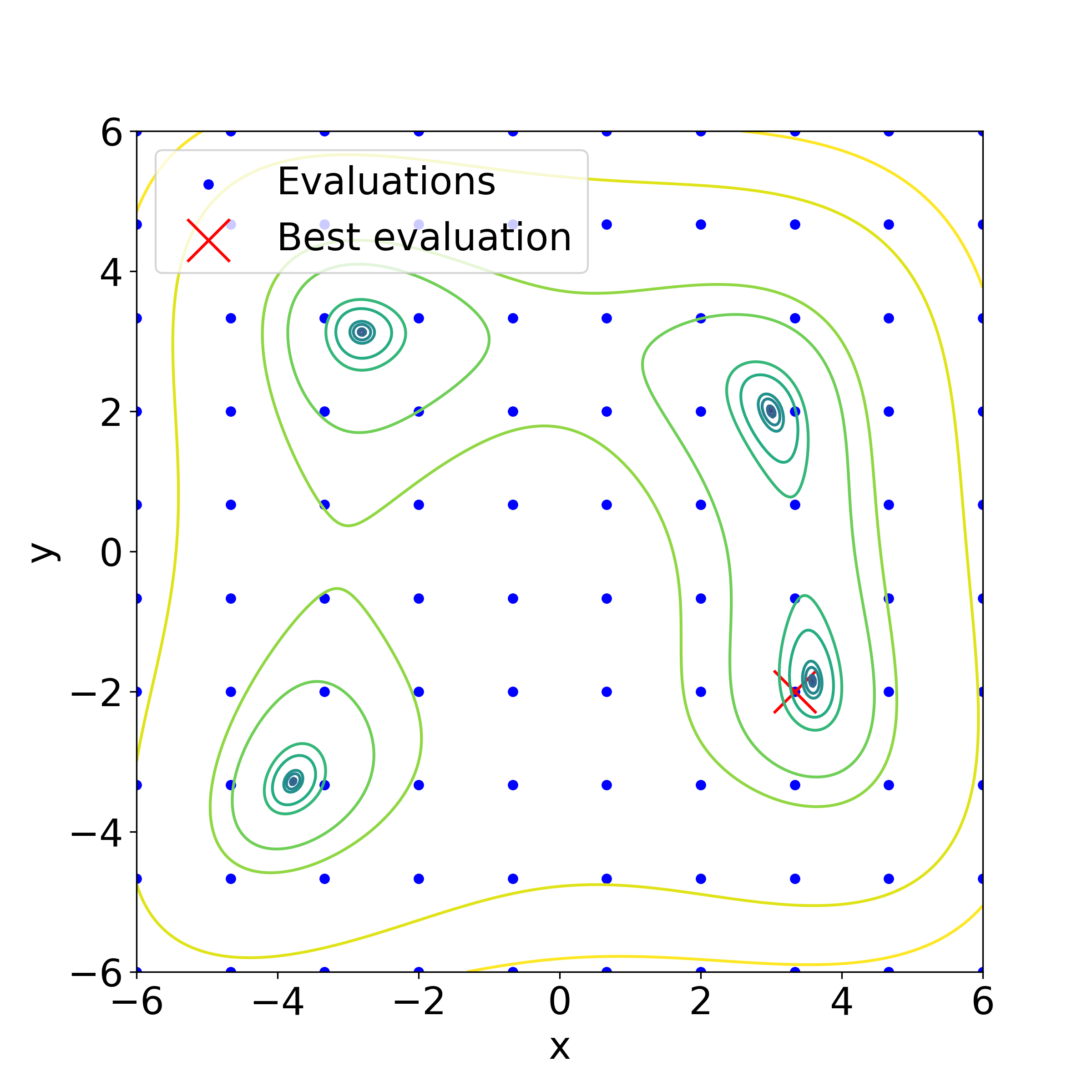}
\label{fig:figs/grid_optimization_himmelblau.png}
\endminipage\hfill
\minipage{0.32\textwidth}
\includegraphics[width=\linewidth]{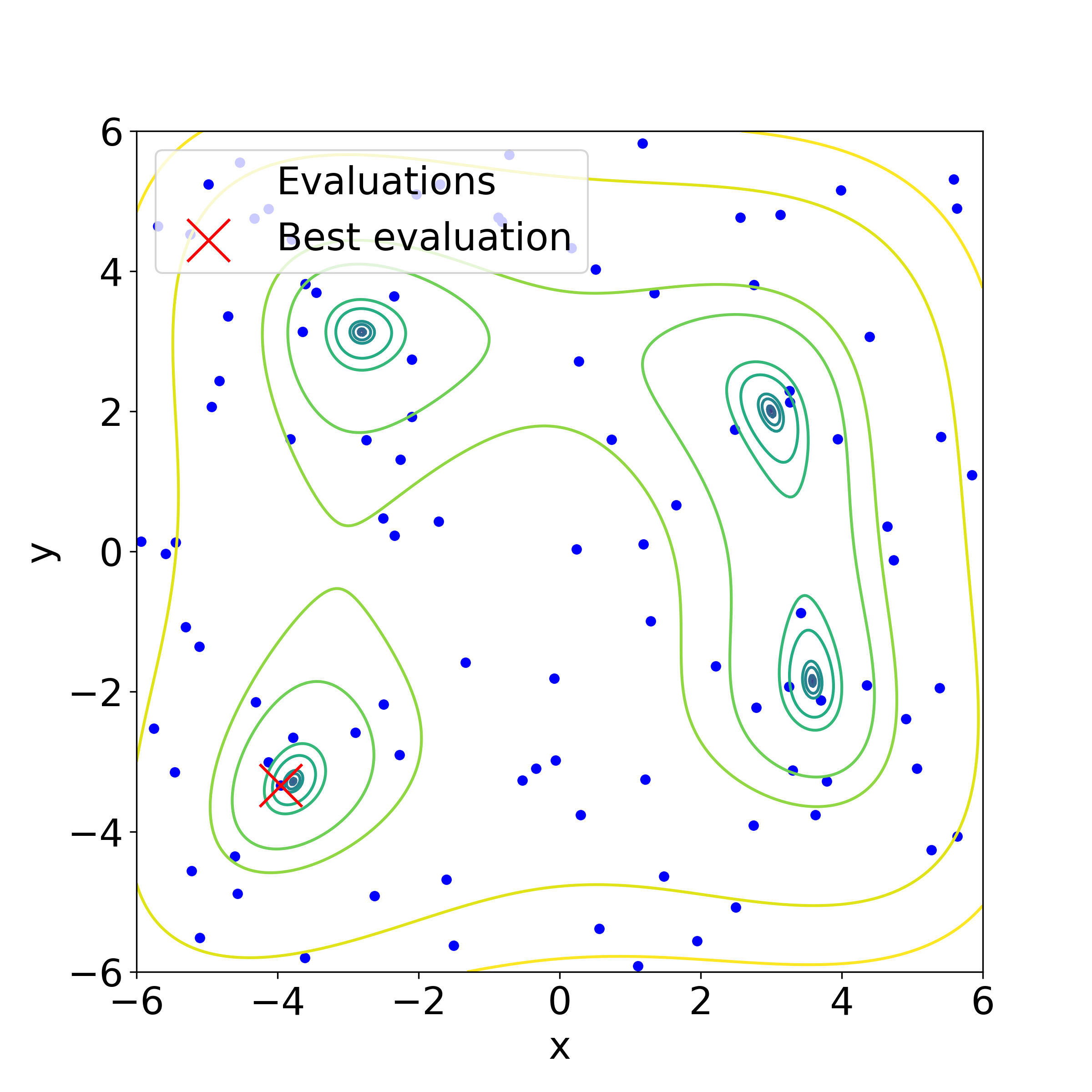}
\label{fig:figs/random_optimization_himmelblau.png}
\endminipage\hfill
\minipage{0.32\textwidth}%
\includegraphics[width=\linewidth]{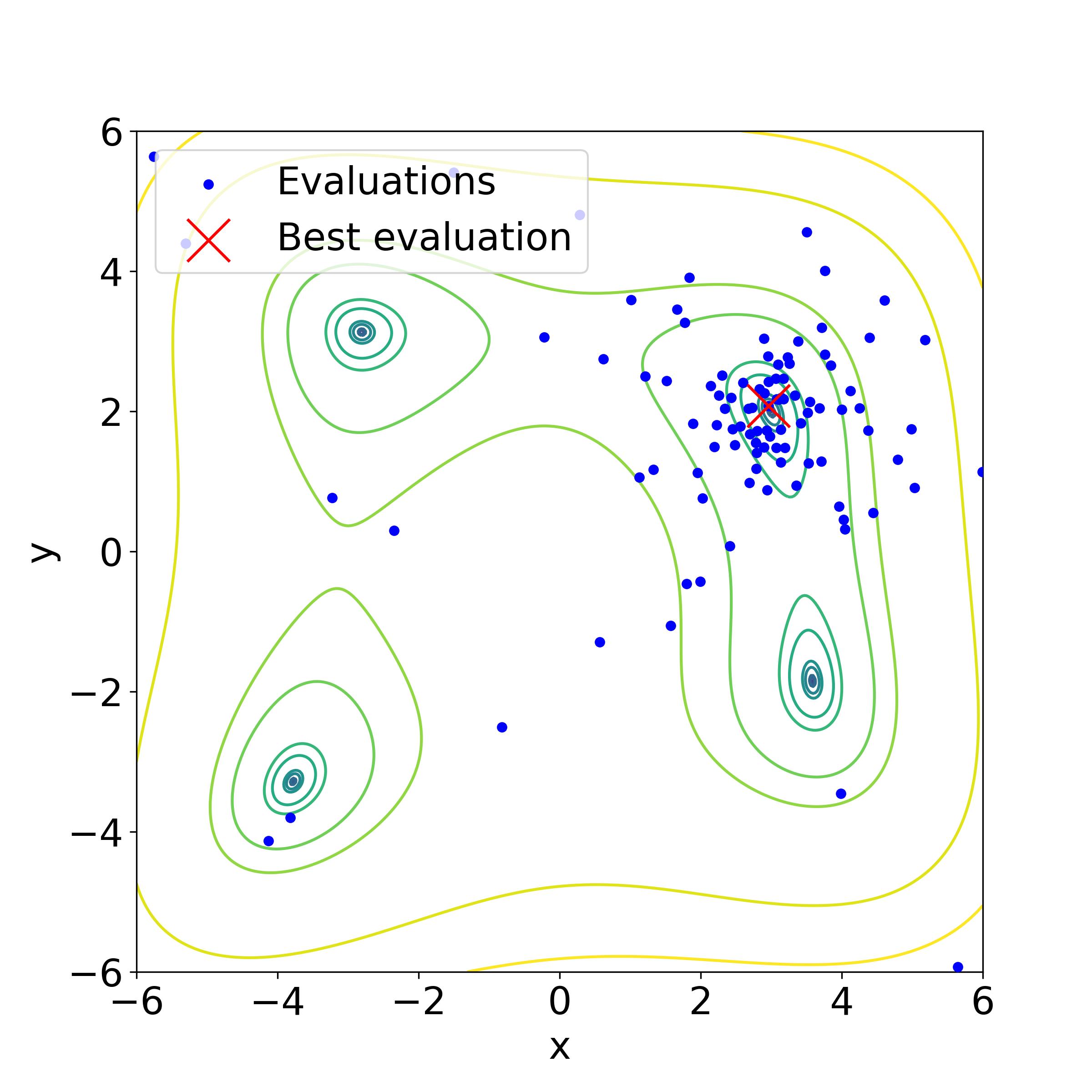}
\endminipage
\caption{Optimizations with grid search (left), random search (center) and Bayesian optimization (right). Contours represent level curves of the Himmelblau function, the blue dots the position of the different evaluations and the red crosses the best trial of each case. As visually seen, Bayesian optimization tends to make trials.}
\label{fig:optimization_himmelblau}
\end{figure}

Figure~\ref{fig:optimization_himmelblau} shows the search patterns resulting from each approach. After $100$ trials, Bayesian optimization found the lowest value among the three methods (recall that the four local minima are at $f=0$), with $f=0.09$, while random search found $f=1.71$ and grid search found $f=3.68$. As can be seen from Fig.~\ref{fig:optimization_himmelblau},  Bayesian optimization has the advantage of exploring the search space adaptively, leading to faster convergence when compared to grid or random searches. This is so because neither grid nor random search keeps evidence from older trials, which makes them less effective than a Bayesian approach for the task.

This example showcases the key features of Bayesian optimization,  
which are: efficiency (i.e., number of trials to reach a minimum), adaptive search, and a relatively low computational effort.

\section{Hyper-Optimized hp-greedy reduced bases for gravitational waves} \label{sec:results}

In this section we apply hp-greedy to build accurate and low dimensional representations of gravitational waves, optimizing the search of hyperparameters with a Bayesian approach, as described in the previous section. 

\subsection{Physical Setup} \label{sec:dataset}
The waveforms used to train hp-greedy and perform HPO were obtained from  NRHybSur3dq8~\cite{varma2019surrogate}. This is a surrogate model for hybridized non-precessing numerical relativity and post-newtonian waveforms within the parameter range of mass ratios $1 \le q \le 8$, and dimensionless spins $-0.8 \le \chi_{1z}, \chi_{2z} \le 0.8$. We focus in this work on the dominant angular mode $l= m = 2$ of the waveforms, which we sampled in the late inspiral and merger phases, with $t \in [-2750, 100]M$, and a time step of $\Delta t = 0.1M$. Additionally, we normalized the waveforms with respect to the $\ell_2$ norm to emphasize structural characteristics rather than size or amplitude.

In this paper we focus on two distinct cases:
\begin{itemize}
    \item 1D Case: This scenario involves no spin, where the sole free parameter is the mass ratio, $q:= m_1/m_2$. 
    \item 2D Case: Two spins aligned in the same direction and with equal magnitudes, $\chi_{1z} = \chi_{2z}$, are added to the 1D scenario.
\end{itemize}

In both cases we generated multiple training sets of different sizes as shown in Figure~\ref{fig:learning_curve}. As error metric for the reduced basis representation $\tilde {h}_{\lambda}(t)$ of a waveform $h_{\lambda}(t)$ labeled by the parameter $\lambda$, we use the maximum in the parameter space of the $\ell_2$ norm, 
\begin{equation}
\epsilon := \max_{\lambda} \| h_{\lambda} (\cdot) - \tilde{h}_{\lambda} (\cdot) \|^{2} \label{eq:error} \, , 
\end{equation}
where $\tilde{h}_{\lambda} (t) :=  \mathcal{P} h_{\lambda} (t) $ is the orthogonal projection of $h_{\lambda}(t)$ onto the span of the basis. For the quadrature involved in the computation of the 2-norm, we use Riemann's rule. 

\subsection{Optimization Methods Compared} \label{sec:methods-compared}
In section \ref{sec:benchmark} we compared the TPE algorithm with random and grid searches. Here we benchmark these methods in the context of gravitational waves using a small dataset in the 1D case. We used 400 waveforms for training and 800 for validation, all equally spaced in the 1D parameter domain ($1 < q < 8$) .

The hyperparameters being optimized were $l_{max}$ (with $ 0 \leq l_{max} \leq 7$) and the seed $\hat{\Lambda}_0$, giving place to a search space with 3,200 different configurations, with $400$ different seeds and 8 possible $l_{max}$ values.
Since we have a discrete search space, we can go through all the different configurations with one grid search. 
We divide the comparison into two parts:

\begin{enumerate}
\item  On the speed of convergence of TPE compared to random search, and how consistent it is through multiple runs.
\item On the time difference between grid search and one run of  the TPE optimization.
\end{enumerate}

\begin{figure}[h!]
\centering
\includegraphics[width=.6\columnwidth]{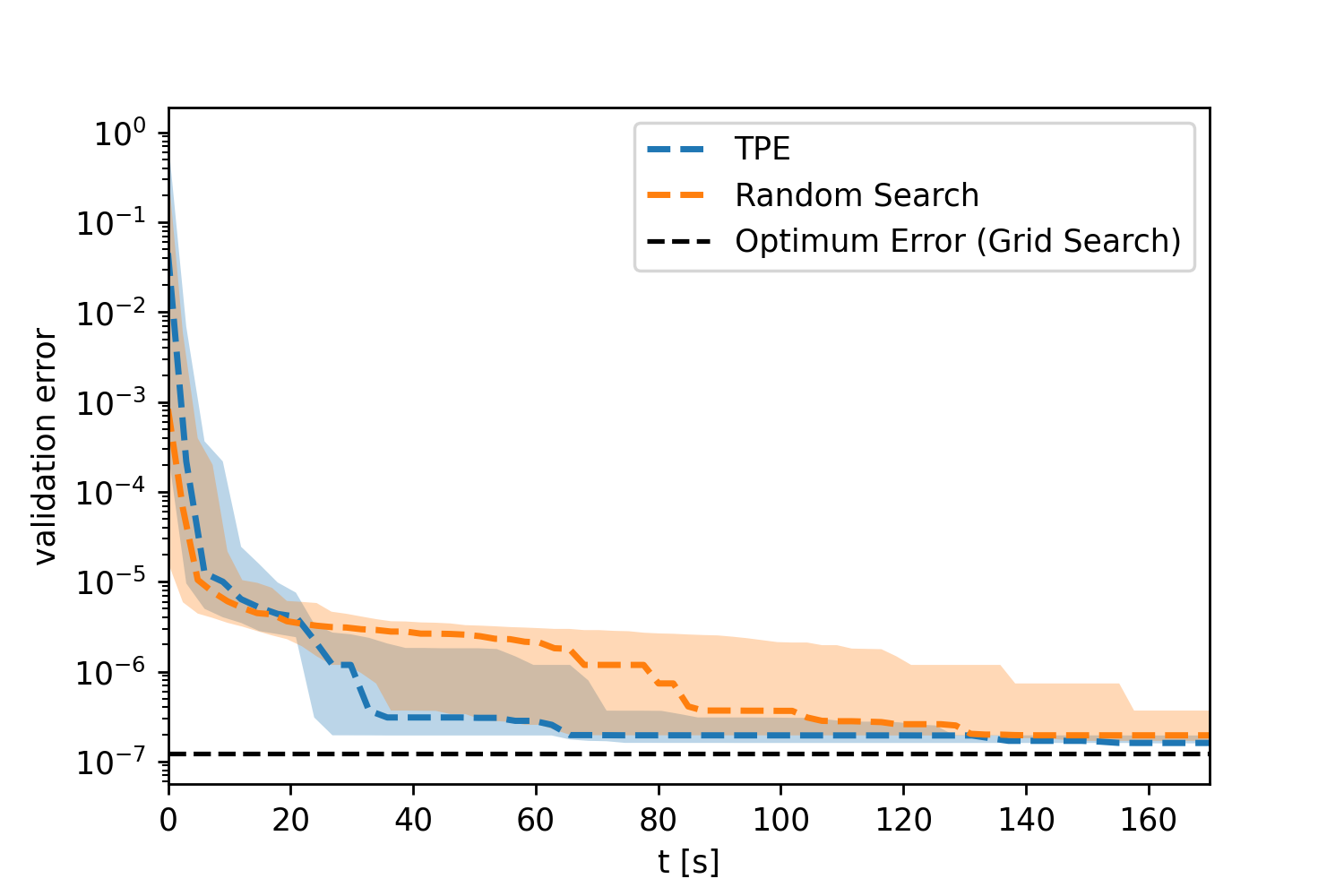}
\caption{Evolution of the best validation error found for TPE and random search with 400 waveforms for training and 800 for validation. The dashed lines represent the median of the best error found for 150 optimizations at a given time, while the shaded area indicates the interquartile range, from Q1 to Q3 (0.25 and 0.75). The black line depicts the optimum error found in the grid search.}
\label{fig:optimization_methods_compared}
\end{figure}

In Figure~ \ref{fig:optimization_methods_compared} we show the results of running 150 optimizations for TPE and random search: each point is the median of the best validation error found for the 150 optimizations at a given time, and the area of color represents the Q1 and Q3 quartiles ($0.25$ and $0.75$, respectively). The black dashed lines show the best error found with grid search. We can see that the TPE curve is always below the random one, and that the colored area reduces drastically around 70 seconds of optimization for the TPE method. The latter shows that TPE consistently finds good results after this point. The optimum value found by grid search was a validation error of $1.23\times 10^{-7}$. This value was found on $19\%$ of the TPE runs, while th other $81\%$ found the second best value $1.59\times 10^{-7}$. On the other hand, none of the random search runs was able to find the optimum value, and only $10\%$ found the second best value of $1.59\times 10^{-7}$. These results show that TPE can find better results than random search, in less time, and more consistently.

Grid search, with the 3200 configurations, took $9.8$ hours to complete, meanwhile 50 iterations of the TPE optimization took about 5 minutes. {\em This is a difference of two orders of magnitude,  a factor of $117 \times$ times in speedup}. In a more realistic optimization task, e.g. using $5,000$ waveforms for training and $10,000$ for validation, and a total of $50,000$ hyperparameter configurations, $100$ iterations of TPE would take around 16 hours. Using these values we can estimate that a grid search would take, in contrast, around $8,000$ hours, or {\em 11 months to complete}. Thus grid search is not a viable method for finding an optimal configuration in any realistic scenario.

All of these runs were performed in the Serafín cluster from CCAD-UNC~\footnote{Full details of the Serafín cluster at \url{https://ccad.unc.edu.ar/equipamiento/cluster-serafin/}.}, where each node consists on 2 AMD EPYC 7532 of 200W with 32 cores of Zen2 microarchitecture, with 128 GiB of RAM DDR4-3200.

\subsection{Optimized hp-greedy reduced bases versus global ones} \label{sec:performance}

\begin{figure}[!h]
\minipage{0.49\textwidth}
\includegraphics[width=\linewidth]{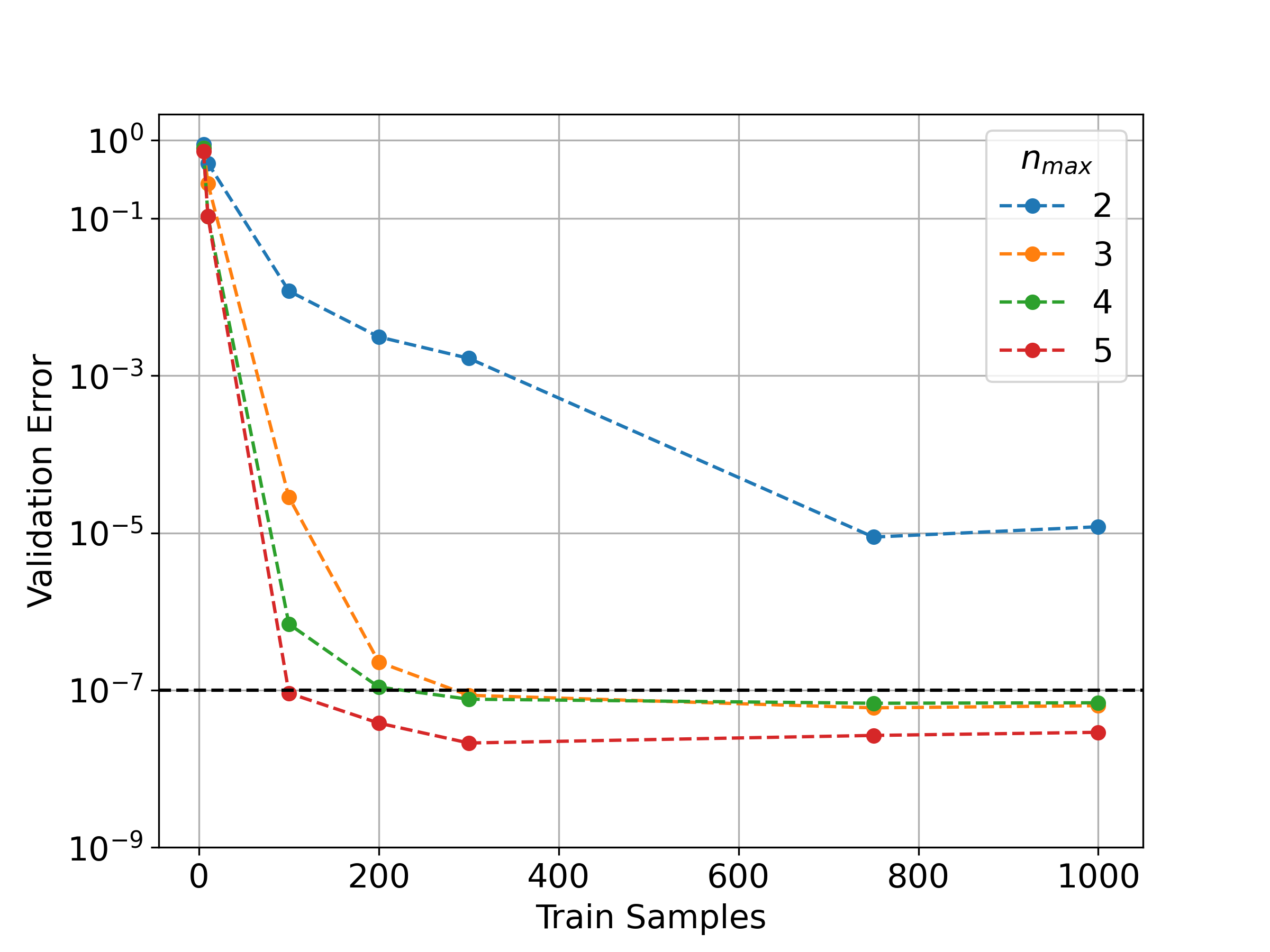}
\label{fig:1d_learning_curve}
\endminipage\hfill
\minipage{0.49\textwidth}
\includegraphics[width=\linewidth]{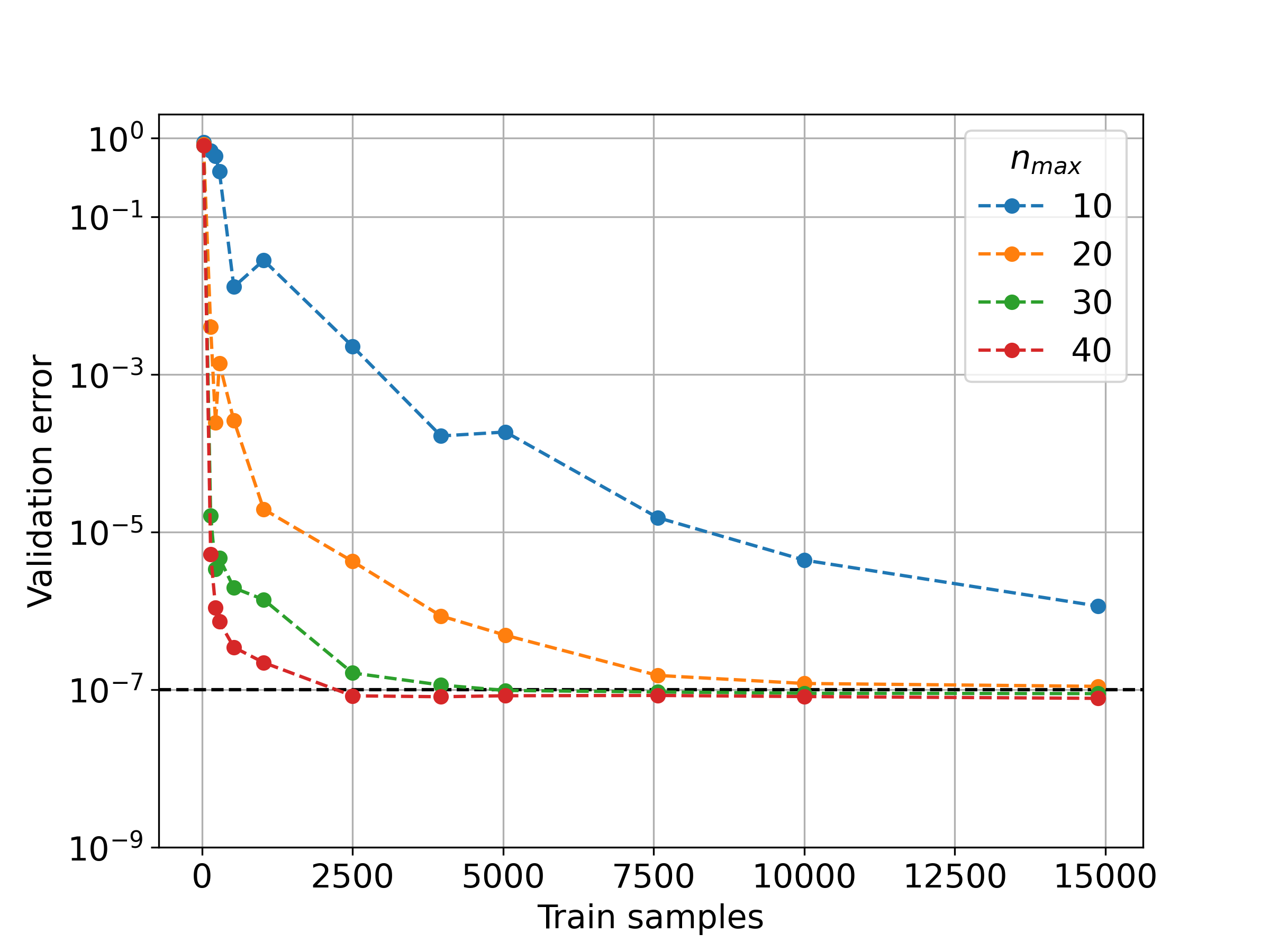}
\label{fig:2d_hpgreedy_learning_curve}
\endminipage\hfill
\caption{Left and right panels: 1D and 2D learning curves for gravitational waves. Each curve represents the validation error of hyperparameter-optimized hp-greedy reduced bases for a fixed $n_{max}$ and varying training samples size. Each dot represents an optimized hp-greedy basis with respect to $l_{max}$ and $\hat{\Lambda}_0$. The dashed horizontal black line represents the value of $\epsilon_{tol} = 10^{-7}$. }
\label{fig:learning_curve}
\end{figure}

Here we present our results for Bayesian HPO and hp-greedy of the gravitational waveforms setup described in Section~\ref{sec:dataset} for the hyperparameters 
$$
\{ \hat{\Lambda}_0 \, , l_{max} \}, 
$$
with fixed maximum dimensionalities $n_{max}$ of each reduced basis. The accuracy threshold in all cases is chosen to be $\epsilon=10^{-7}$, which is the systematic error of the underlying model, NRHybSur3dq8~\cite{varma2019surrogate}. 

The {\it learning curves} of figure [\ref{fig:learning_curve}] show the validation error achieved for each optimized hp-greedy reduced basis; the intent of these plots is to determine when a  training set is dense enough. For example, in the 1D case around $2,000$ training samples are enough; while in the 2D case this number grows to $\sim 15,000$, which is smaller than $2,000 \times 2,000$ and shows that there is increased redundancy in the waveforms as the dimensionality grows. 

\begin{figure}[!h]
\minipage{0.49\textwidth}
\includegraphics[width=\linewidth]{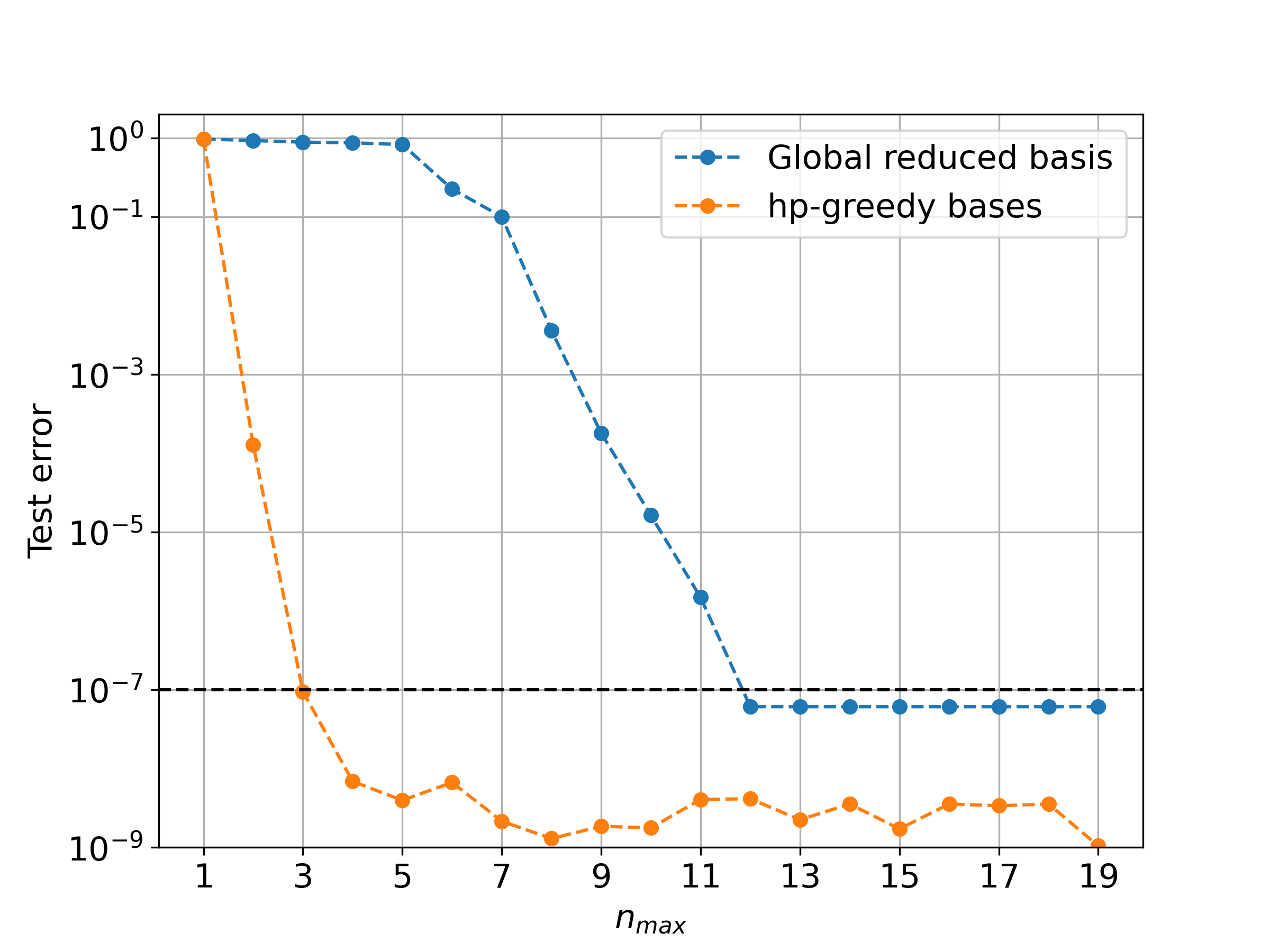}
\label{fig:1d_test}
\endminipage\hfill
\minipage{0.49\textwidth}
\includegraphics[width=\linewidth]{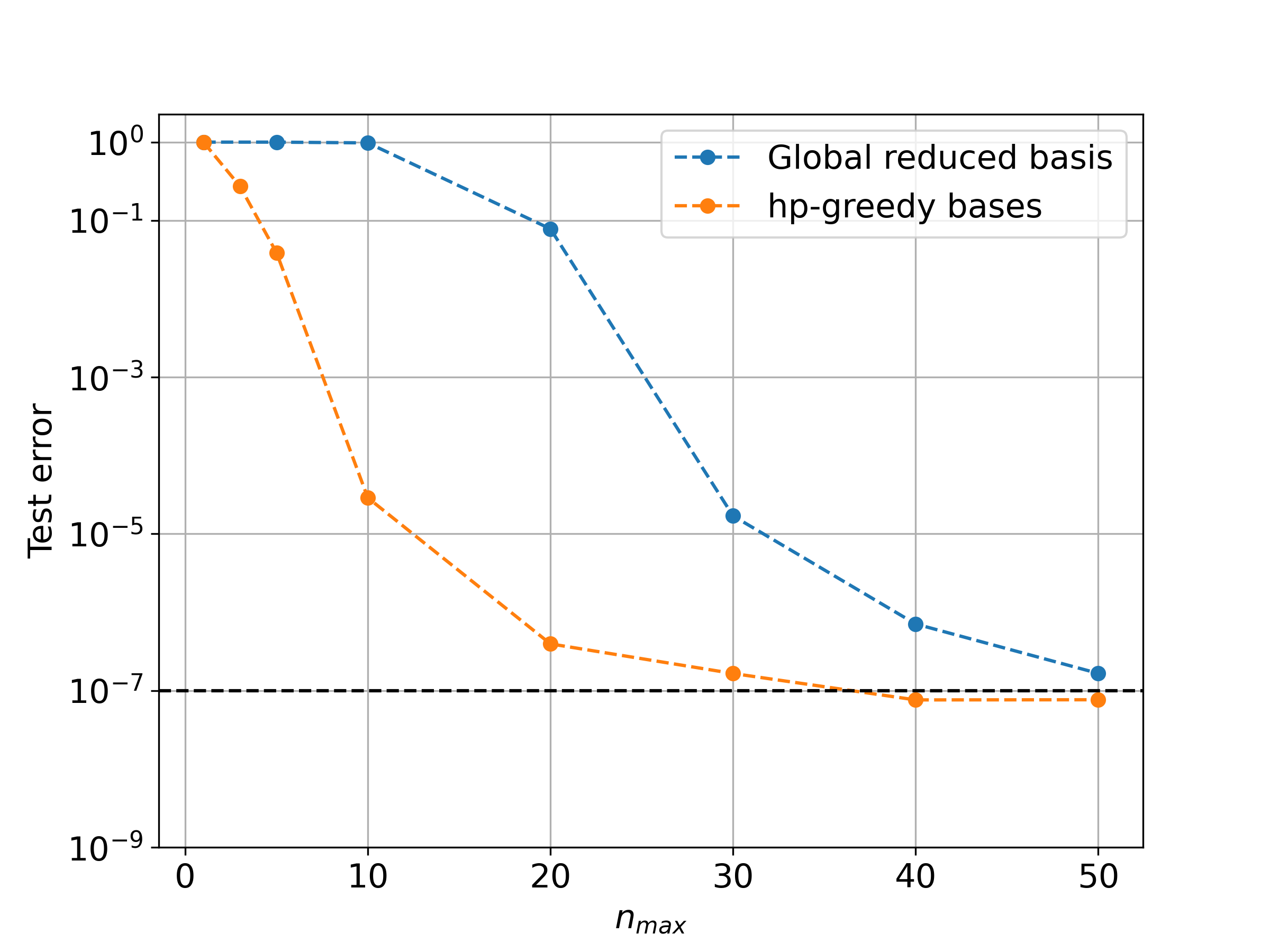}
\label{fig:2d_test}
\endminipage\hfill
\caption{Test errors comparing global reduced basis with local, hp-greedy ones.}
\label{fig:test}
\end{figure}

We are interested in the smallest $n_{max}$ for each case, since this implies the fastest online waveform evaluation and data analysis in general: these are $n_{max}=3, 10$ for the 1D and 2D cases, respectively. When compared to global bases, the hyperparameter-optimized hp-greedy bases are $4-5$ times smaller, which should translate into a factor of $(4-5)\times$ speedup both in waveform evaluation but, most importantly, in parameter estimation. 

\section{Discussion} \label{sec:com}
In this paper we continued our work on local, unstructured reduced bases for gravitational waves using hp-greedy refinement, with the aim of accelerating both waveform predictions and inference (especially parameter estimation). In reference~\cite{Cerino:2022dhr} we found that there are new hyperparameters to be optimized, which do not appear in global reduced basis.  As usual in the ML context, parameters are learned from training, while hyperparameters remain fixed during each training stage. 

The resulting structure of hp-greedy reduced bases is that of a binary tree. In our simulations, though limited in size, we have empirically found that the trees of hp-greedy refinement end up being nearly balanced. When a representation is needed for a given parameter, the corresponding local basis can be searched for in the representation tree in an efficient way, avoiding the computational cost of a direct/naive search.  To do so, two sequential steps are needed: i) find the subspace containing the local reduced basis, and ii) use that basis for the representation. The search utilizes $\lambda$ as input and the hp-greedy tree structure to traverse the tree from the root to a leaf node, which contains the queried value of $\lambda$ in its subspace (see equation $(4.13)$ of \cite{EftangThesis} for more details). The advantage of this approach is the low computational cost to find the required subspace; for example, if there are $n$ subspaces and the tree is balanced, the computational cost is of order $\mathcal{O}$(log $n$).

There are several stopping criteria for subdividing the parameter domain and avoiding overfitting. In this work we have taken them to be $n_{max}$ (maximum dimensionality of each local reduced basis), $l_{max}$ (maximum depth of the tree) and the error $\epsilon_{tol}$; in practice the latter should not be smaller than the underlying systematical error of the data.  

Our results show that using a Bayesian approach is a promising path to HPO. Nonetheless, there are other alternatives such as evolutionary programming or genetic algorithms, which were left out of the scope of this paper and should be analyzed in future work. 

In conjunction with the computations outlined in the paper, we have made the corresponding code available on GitHub. These repositories contain the codebase for Bayesian optimization \cite{hp-greedy-bayesian-optimization} and hp-greedy models \cite{scikit-reducedmodel} used in the paper.

\section{Acknowledgments} \label{sec:ack}
This work was partially supported by CONICET and by project PICT-2021-00757, Argentina, it used computational resources from CCAD – Universidad Nacional de Córdoba (https://ccad.unc.edu.ar/), which are part of SNCAD – MinCyT, República Argentina. MT thanks the Horace Hearne Institute for Theoretical Physics at LSU for hospitality during the conference ``Workshop on Gravity: classical, quantum, theoretical and experimental'' on March 2023, where part of this work was done. 

\bibliography{hpo}

\begin{thebibliography}{60}%
\makeatletter
\providecommand \@ifxundefined [1]{%
 \@ifx{#1\undefined}
}%
\providecommand \@ifnum [1]{%
 \ifnum #1\expandafter \@firstoftwo
 \else \expandafter \@secondoftwo
 \fi
}%
\providecommand \@ifx [1]{%
 \ifx #1\expandafter \@firstoftwo
 \else \expandafter \@secondoftwo
 \fi
}%
\providecommand \natexlab [1]{#1}%
\providecommand \enquote  [1]{``#1''}%
\providecommand \bibnamefont  [1]{#1}%
\providecommand \bibfnamefont [1]{#1}%
\providecommand \citenamefont [1]{#1}%
\providecommand \href@noop [0]{\@secondoftwo}%
\providecommand \href [0]{\begingroup \@sanitize@url \@href}%
\providecommand \@href[1]{\@@startlink{#1}\@@href}%
\providecommand \@@href[1]{\endgroup#1\@@endlink}%
\providecommand \@sanitize@url [0]{\catcode `\\12\catcode `\$12\catcode `\&12\catcode `\#12\catcode `\^12\catcode `\_12\catcode `\%12\relax}%
\providecommand \@@startlink[1]{}%
\providecommand \@@endlink[0]{}%
\providecommand \url  [0]{\begingroup\@sanitize@url \@url }%
\providecommand \@url [1]{\endgroup\@href {#1}{\urlprefix }}%
\providecommand \urlprefix  [0]{URL }%
\providecommand \Eprint [0]{\href }%
\providecommand \doibase [0]{https://doi.org/}%
\providecommand \selectlanguage [0]{\@gobble}%
\providecommand \bibinfo  [0]{\@secondoftwo}%
\providecommand \bibfield  [0]{\@secondoftwo}%
\providecommand \translation [1]{[#1]}%
\providecommand \BibitemOpen [0]{}%
\providecommand \bibitemStop [0]{}%
\providecommand \bibitemNoStop [0]{.\EOS\space}%
\providecommand \EOS [0]{\spacefactor3000\relax}%
\providecommand \BibitemShut  [1]{\csname bibitem#1\endcsname}%
\let\auto@bib@innerbib\@empty
\bibitem [{\citenamefont {Cerino}\ \emph {et~al.}(2023)\citenamefont {Cerino}, \citenamefont {Diaz-Pace},\ and\ \citenamefont {Tiglio}}]{Cerino:2022dhr}%
  \BibitemOpen
  \bibfield  {author} {\bibinfo {author} {\bibfnamefont {F.}~\bibnamefont {Cerino}}, \bibinfo {author} {\bibfnamefont {J.~A.}\ \bibnamefont {Diaz-Pace}},\ and\ \bibinfo {author} {\bibfnamefont {M.}~\bibnamefont {Tiglio}},\ }\bibfield  {title} {\bibinfo {title} {{An automated parameter domain decomposition approach for gravitational wave surrogates using hp-greedy refinement}},\ }\href {https://doi.org/10.1088/1361-6382/acf4e7} {\bibfield  {journal} {\bibinfo  {journal} {Class. Quant. Grav.}\ }\textbf {\bibinfo {volume} {40}},\ \bibinfo {pages} {205003} (\bibinfo {year} {2023})},\ \Eprint {https://arxiv.org/abs/2212.08554} {arXiv:2212.08554 [gr-qc]} \BibitemShut {NoStop}%
\bibitem [{\citenamefont {Tiglio}\ and\ \citenamefont {Villanueva}(2022)}]{tiglio2022reduced}%
  \BibitemOpen
  \bibfield  {author} {\bibinfo {author} {\bibfnamefont {M.}~\bibnamefont {Tiglio}}\ and\ \bibinfo {author} {\bibfnamefont {A.}~\bibnamefont {Villanueva}},\ }\bibfield  {title} {\bibinfo {title} {Reduced order and surrogate models for gravitational waves},\ }\href@noop {} {\bibfield  {journal} {\bibinfo  {journal} {Living Reviews in Relativity}\ }\textbf {\bibinfo {volume} {25}},\ \bibinfo {pages} {2} (\bibinfo {year} {2022})}\BibitemShut {NoStop}%
\bibitem [{\citenamefont {Mandic}\ \emph {et~al.}(2012)\citenamefont {Mandic}, \citenamefont {Thrane}, \citenamefont {Giampanis},\ and\ \citenamefont {Regimbau}}]{mandic2012parameter}%
  \BibitemOpen
  \bibfield  {author} {\bibinfo {author} {\bibfnamefont {V.}~\bibnamefont {Mandic}}, \bibinfo {author} {\bibfnamefont {E.}~\bibnamefont {Thrane}}, \bibinfo {author} {\bibfnamefont {S.}~\bibnamefont {Giampanis}},\ and\ \bibinfo {author} {\bibfnamefont {T.}~\bibnamefont {Regimbau}},\ }\bibfield  {title} {\bibinfo {title} {Parameter estimation in searches for the stochastic gravitational-wave background},\ }\href@noop {} {\bibfield  {journal} {\bibinfo  {journal} {Physical Review Letters}\ }\textbf {\bibinfo {volume} {109}},\ \bibinfo {pages} {171102} (\bibinfo {year} {2012})}\BibitemShut {NoStop}%
\bibitem [{\citenamefont {Isi}\ \emph {et~al.}(2019)\citenamefont {Isi}, \citenamefont {Chatziioannou},\ and\ \citenamefont {Farr}}]{isi2019hierarchical}%
  \BibitemOpen
  \bibfield  {author} {\bibinfo {author} {\bibfnamefont {M.}~\bibnamefont {Isi}}, \bibinfo {author} {\bibfnamefont {K.}~\bibnamefont {Chatziioannou}},\ and\ \bibinfo {author} {\bibfnamefont {W.~M.}\ \bibnamefont {Farr}},\ }\bibfield  {title} {\bibinfo {title} {Hierarchical test of general relativity with gravitational waves},\ }\href@noop {} {\bibfield  {journal} {\bibinfo  {journal} {Physical Review Letters}\ }\textbf {\bibinfo {volume} {123}},\ \bibinfo {pages} {121101} (\bibinfo {year} {2019})}\BibitemShut {NoStop}%
\bibitem [{\citenamefont {Lange}\ \emph {et~al.}(2018)\citenamefont {Lange}, \citenamefont {O'Shaughnessy},\ and\ \citenamefont {Rizzo}}]{lange2018rapid}%
  \BibitemOpen
  \bibfield  {author} {\bibinfo {author} {\bibfnamefont {J.}~\bibnamefont {Lange}}, \bibinfo {author} {\bibfnamefont {R.}~\bibnamefont {O'Shaughnessy}},\ and\ \bibinfo {author} {\bibfnamefont {M.}~\bibnamefont {Rizzo}},\ }\bibfield  {title} {\bibinfo {title} {Rapid and accurate parameter inference for coalescing, precessing compact binaries},\ }\href@noop {} {\bibfield  {journal} {\bibinfo  {journal} {arXiv preprint arXiv:1805.10457}\ } (\bibinfo {year} {2018})}\BibitemShut {NoStop}%
\bibitem [{\citenamefont {Lynch}\ \emph {et~al.}(2017)\citenamefont {Lynch}, \citenamefont {Vitale}, \citenamefont {Essick}, \citenamefont {Katsavounidis},\ and\ \citenamefont {Robinet}}]{lynch2017information}%
  \BibitemOpen
  \bibfield  {author} {\bibinfo {author} {\bibfnamefont {R.}~\bibnamefont {Lynch}}, \bibinfo {author} {\bibfnamefont {S.}~\bibnamefont {Vitale}}, \bibinfo {author} {\bibfnamefont {R.}~\bibnamefont {Essick}}, \bibinfo {author} {\bibfnamefont {E.}~\bibnamefont {Katsavounidis}},\ and\ \bibinfo {author} {\bibfnamefont {F.}~\bibnamefont {Robinet}},\ }\bibfield  {title} {\bibinfo {title} {Information-theoretic approach to the gravitational-wave burst detection problem},\ }\href@noop {} {\bibfield  {journal} {\bibinfo  {journal} {Physical Review D}\ }\textbf {\bibinfo {volume} {95}},\ \bibinfo {pages} {104046} (\bibinfo {year} {2017})}\BibitemShut {NoStop}%
\bibitem [{\citenamefont {Mandel}\ \emph {et~al.}(2014)\citenamefont {Mandel}, \citenamefont {Berry}, \citenamefont {Ohme}, \citenamefont {Fairhurst},\ and\ \citenamefont {Farr}}]{mandel2014parameter}%
  \BibitemOpen
  \bibfield  {author} {\bibinfo {author} {\bibfnamefont {I.}~\bibnamefont {Mandel}}, \bibinfo {author} {\bibfnamefont {C.~P.}\ \bibnamefont {Berry}}, \bibinfo {author} {\bibfnamefont {F.}~\bibnamefont {Ohme}}, \bibinfo {author} {\bibfnamefont {S.}~\bibnamefont {Fairhurst}},\ and\ \bibinfo {author} {\bibfnamefont {W.~M.}\ \bibnamefont {Farr}},\ }\bibfield  {title} {\bibinfo {title} {Parameter estimation on compact binary coalescences with abruptly terminating gravitational waveforms},\ }\href@noop {} {\bibfield  {journal} {\bibinfo  {journal} {Classical and Quantum Gravity}\ }\textbf {\bibinfo {volume} {31}},\ \bibinfo {pages} {155005} (\bibinfo {year} {2014})}\BibitemShut {NoStop}%
\bibitem [{\citenamefont {Mandel}\ \emph {et~al.}(2019)\citenamefont {Mandel}, \citenamefont {Farr},\ and\ \citenamefont {Gair}}]{mandel2019extracting}%
  \BibitemOpen
  \bibfield  {author} {\bibinfo {author} {\bibfnamefont {I.}~\bibnamefont {Mandel}}, \bibinfo {author} {\bibfnamefont {W.~M.}\ \bibnamefont {Farr}},\ and\ \bibinfo {author} {\bibfnamefont {J.~R.}\ \bibnamefont {Gair}},\ }\bibfield  {title} {\bibinfo {title} {Extracting distribution parameters from multiple uncertain observations with selection biases},\ }\href@noop {} {\bibfield  {journal} {\bibinfo  {journal} {Monthly Notices of the Royal Astronomical Society}\ }\textbf {\bibinfo {volume} {486}},\ \bibinfo {pages} {1086} (\bibinfo {year} {2019})}\BibitemShut {NoStop}%
\bibitem [{\citenamefont {Usman}\ \emph {et~al.}(2019)\citenamefont {Usman}, \citenamefont {Mills},\ and\ \citenamefont {Fairhurst}}]{usman2019constraining}%
  \BibitemOpen
  \bibfield  {author} {\bibinfo {author} {\bibfnamefont {S.~A.}\ \bibnamefont {Usman}}, \bibinfo {author} {\bibfnamefont {J.~C.}\ \bibnamefont {Mills}},\ and\ \bibinfo {author} {\bibfnamefont {S.}~\bibnamefont {Fairhurst}},\ }\bibfield  {title} {\bibinfo {title} {Constraining the inclinations of binary mergers from gravitational-wave observations},\ }\href@noop {} {\bibfield  {journal} {\bibinfo  {journal} {The Astrophysical Journal}\ }\textbf {\bibinfo {volume} {877}},\ \bibinfo {pages} {82} (\bibinfo {year} {2019})}\BibitemShut {NoStop}%
\bibitem [{\citenamefont {Van Der~Sluys}\ \emph {et~al.}(2009)\citenamefont {Van Der~Sluys}, \citenamefont {Mandel}, \citenamefont {Raymond}, \citenamefont {Kalogera}, \citenamefont {R{\"o}ver},\ and\ \citenamefont {Christensen}}]{van2009parameter}%
  \BibitemOpen
  \bibfield  {author} {\bibinfo {author} {\bibfnamefont {M.}~\bibnamefont {Van Der~Sluys}}, \bibinfo {author} {\bibfnamefont {I.}~\bibnamefont {Mandel}}, \bibinfo {author} {\bibfnamefont {V.}~\bibnamefont {Raymond}}, \bibinfo {author} {\bibfnamefont {V.}~\bibnamefont {Kalogera}}, \bibinfo {author} {\bibfnamefont {C.}~\bibnamefont {R{\"o}ver}},\ and\ \bibinfo {author} {\bibfnamefont {N.}~\bibnamefont {Christensen}},\ }\bibfield  {title} {\bibinfo {title} {Parameter estimation for signals from compact binary inspirals injected into ligo data},\ }\href@noop {} {\bibfield  {journal} {\bibinfo  {journal} {Classical and Quantum Gravity}\ }\textbf {\bibinfo {volume} {26}},\ \bibinfo {pages} {204010} (\bibinfo {year} {2009})}\BibitemShut {NoStop}%
\bibitem [{\citenamefont {Fishbach}\ \emph {et~al.}(2020)\citenamefont {Fishbach}, \citenamefont {Essick},\ and\ \citenamefont {Holz}}]{fishbach2020does}%
  \BibitemOpen
  \bibfield  {author} {\bibinfo {author} {\bibfnamefont {M.}~\bibnamefont {Fishbach}}, \bibinfo {author} {\bibfnamefont {R.}~\bibnamefont {Essick}},\ and\ \bibinfo {author} {\bibfnamefont {D.~E.}\ \bibnamefont {Holz}},\ }\bibfield  {title} {\bibinfo {title} {Does matter matter? using the mass distribution to distinguish neutron stars and black holes},\ }\href@noop {} {\bibfield  {journal} {\bibinfo  {journal} {The Astrophysical Journal Letters}\ }\textbf {\bibinfo {volume} {899}},\ \bibinfo {pages} {L8} (\bibinfo {year} {2020})}\BibitemShut {NoStop}%
\bibitem [{\citenamefont {Cornish}(2021)}]{cornish2021rapid}%
  \BibitemOpen
  \bibfield  {author} {\bibinfo {author} {\bibfnamefont {N.~J.}\ \bibnamefont {Cornish}},\ }\bibfield  {title} {\bibinfo {title} {Rapid and robust parameter inference for binary mergers},\ }\href@noop {} {\bibfield  {journal} {\bibinfo  {journal} {Physical Review D}\ }\textbf {\bibinfo {volume} {103}},\ \bibinfo {pages} {104057} (\bibinfo {year} {2021})}\BibitemShut {NoStop}%
\bibitem [{\citenamefont {Berry}\ \emph {et~al.}(2015)\citenamefont {Berry}, \citenamefont {Mandel}, \citenamefont {Middleton}, \citenamefont {Singer}, \citenamefont {Urban}, \citenamefont {Vecchio}, \citenamefont {Vitale}, \citenamefont {Cannon}, \citenamefont {Farr}, \citenamefont {Farr} \emph {et~al.}}]{berry2015parameter}%
  \BibitemOpen
  \bibfield  {author} {\bibinfo {author} {\bibfnamefont {C.~P.}\ \bibnamefont {Berry}}, \bibinfo {author} {\bibfnamefont {I.}~\bibnamefont {Mandel}}, \bibinfo {author} {\bibfnamefont {H.}~\bibnamefont {Middleton}}, \bibinfo {author} {\bibfnamefont {L.~P.}\ \bibnamefont {Singer}}, \bibinfo {author} {\bibfnamefont {A.~L.}\ \bibnamefont {Urban}}, \bibinfo {author} {\bibfnamefont {A.}~\bibnamefont {Vecchio}}, \bibinfo {author} {\bibfnamefont {S.}~\bibnamefont {Vitale}}, \bibinfo {author} {\bibfnamefont {K.}~\bibnamefont {Cannon}}, \bibinfo {author} {\bibfnamefont {B.}~\bibnamefont {Farr}}, \bibinfo {author} {\bibfnamefont {W.~M.}\ \bibnamefont {Farr}}, \emph {et~al.},\ }\bibfield  {title} {\bibinfo {title} {Parameter estimation for binary neutron-star coalescences with realistic noise during the advanced ligo era},\ }\href@noop {} {\bibfield  {journal} {\bibinfo  {journal} {The Astrophysical Journal}\ }\textbf {\bibinfo {volume} {804}},\ \bibinfo {pages} {114} (\bibinfo {year} {2015})}\BibitemShut {NoStop}%
\bibitem [{\citenamefont {Biscoveanu}\ \emph {et~al.}(2020)\citenamefont {Biscoveanu}, \citenamefont {Haster}, \citenamefont {Vitale},\ and\ \citenamefont {Davies}}]{biscoveanu2020quantifying}%
  \BibitemOpen
  \bibfield  {author} {\bibinfo {author} {\bibfnamefont {S.}~\bibnamefont {Biscoveanu}}, \bibinfo {author} {\bibfnamefont {C.-J.}\ \bibnamefont {Haster}}, \bibinfo {author} {\bibfnamefont {S.}~\bibnamefont {Vitale}},\ and\ \bibinfo {author} {\bibfnamefont {J.}~\bibnamefont {Davies}},\ }\bibfield  {title} {\bibinfo {title} {Quantifying the effect of power spectral density uncertainty on gravitational-wave parameter estimation for compact binary sources},\ }\href@noop {} {\bibfield  {journal} {\bibinfo  {journal} {Physical Review D}\ }\textbf {\bibinfo {volume} {102}},\ \bibinfo {pages} {023008} (\bibinfo {year} {2020})}\BibitemShut {NoStop}%
\bibitem [{\citenamefont {Bizouard}\ \emph {et~al.}(2021)\citenamefont {Bizouard}, \citenamefont {Maturana-Russel}, \citenamefont {Torres-Forn{\'e}}, \citenamefont {Obergaulinger}, \citenamefont {Cerd{\'a}-Dur{\'a}n}, \citenamefont {Christensen}, \citenamefont {Font},\ and\ \citenamefont {Meyer}}]{bizouard2021inference}%
  \BibitemOpen
  \bibfield  {author} {\bibinfo {author} {\bibfnamefont {M.-A.}\ \bibnamefont {Bizouard}}, \bibinfo {author} {\bibfnamefont {P.}~\bibnamefont {Maturana-Russel}}, \bibinfo {author} {\bibfnamefont {A.}~\bibnamefont {Torres-Forn{\'e}}}, \bibinfo {author} {\bibfnamefont {M.}~\bibnamefont {Obergaulinger}}, \bibinfo {author} {\bibfnamefont {P.}~\bibnamefont {Cerd{\'a}-Dur{\'a}n}}, \bibinfo {author} {\bibfnamefont {N.}~\bibnamefont {Christensen}}, \bibinfo {author} {\bibfnamefont {J.~A.}\ \bibnamefont {Font}},\ and\ \bibinfo {author} {\bibfnamefont {R.}~\bibnamefont {Meyer}},\ }\bibfield  {title} {\bibinfo {title} {Inference of protoneutron star properties from gravitational-wave data in core-collapse supernovae},\ }\href@noop {} {\bibfield  {journal} {\bibinfo  {journal} {Physical Review D}\ }\textbf {\bibinfo {volume} {103}},\ \bibinfo {pages} {063006} (\bibinfo {year} {2021})}\BibitemShut {NoStop}%
\bibitem [{\citenamefont {Banagiri}\ \emph {et~al.}(2020)\citenamefont {Banagiri}, \citenamefont {Coughlin}, \citenamefont {Clark}, \citenamefont {Lasky}, \citenamefont {Bizouard}, \citenamefont {Talbot}, \citenamefont {Thrane},\ and\ \citenamefont {Mandic}}]{banagiri2020constraining}%
  \BibitemOpen
  \bibfield  {author} {\bibinfo {author} {\bibfnamefont {S.}~\bibnamefont {Banagiri}}, \bibinfo {author} {\bibfnamefont {M.~W.}\ \bibnamefont {Coughlin}}, \bibinfo {author} {\bibfnamefont {J.}~\bibnamefont {Clark}}, \bibinfo {author} {\bibfnamefont {P.~D.}\ \bibnamefont {Lasky}}, \bibinfo {author} {\bibfnamefont {M.-A.}\ \bibnamefont {Bizouard}}, \bibinfo {author} {\bibfnamefont {C.}~\bibnamefont {Talbot}}, \bibinfo {author} {\bibfnamefont {E.}~\bibnamefont {Thrane}},\ and\ \bibinfo {author} {\bibfnamefont {V.}~\bibnamefont {Mandic}},\ }\bibfield  {title} {\bibinfo {title} {Constraining the gravitational-wave afterglow from a binary neutron star coalescence},\ }\href@noop {} {\bibfield  {journal} {\bibinfo  {journal} {Monthly Notices of the Royal Astronomical Society}\ }\textbf {\bibinfo {volume} {492}},\ \bibinfo {pages} {4945} (\bibinfo {year} {2020})}\BibitemShut {NoStop}%
\bibitem [{\citenamefont {Coughlin}\ \emph {et~al.}(2019)\citenamefont {Coughlin}, \citenamefont {Dietrich}, \citenamefont {Margalit},\ and\ \citenamefont {Metzger}}]{coughlin2019multimessenger}%
  \BibitemOpen
  \bibfield  {author} {\bibinfo {author} {\bibfnamefont {M.~W.}\ \bibnamefont {Coughlin}}, \bibinfo {author} {\bibfnamefont {T.}~\bibnamefont {Dietrich}}, \bibinfo {author} {\bibfnamefont {B.}~\bibnamefont {Margalit}},\ and\ \bibinfo {author} {\bibfnamefont {B.~D.}\ \bibnamefont {Metzger}},\ }\bibfield  {title} {\bibinfo {title} {Multimessenger bayesian parameter inference of a binary neutron star merger},\ }\href@noop {} {\bibfield  {journal} {\bibinfo  {journal} {Monthly Notices of the Royal Astronomical Society: Letters}\ }\textbf {\bibinfo {volume} {489}},\ \bibinfo {pages} {L91} (\bibinfo {year} {2019})}\BibitemShut {NoStop}%
\bibitem [{\citenamefont {Wysocki}\ \emph {et~al.}(2019)\citenamefont {Wysocki}, \citenamefont {Lange},\ and\ \citenamefont {O’Shaughnessy}}]{wysocki2019reconstructing}%
  \BibitemOpen
  \bibfield  {author} {\bibinfo {author} {\bibfnamefont {D.}~\bibnamefont {Wysocki}}, \bibinfo {author} {\bibfnamefont {J.}~\bibnamefont {Lange}},\ and\ \bibinfo {author} {\bibfnamefont {R.}~\bibnamefont {O’Shaughnessy}},\ }\bibfield  {title} {\bibinfo {title} {Reconstructing phenomenological distributions of compact binaries via gravitational wave observations},\ }\href@noop {} {\bibfield  {journal} {\bibinfo  {journal} {Physical Review D}\ }\textbf {\bibinfo {volume} {100}},\ \bibinfo {pages} {043012} (\bibinfo {year} {2019})}\BibitemShut {NoStop}%
\bibitem [{\citenamefont {Christensen}\ and\ \citenamefont {Meyer}(2022)}]{christensen2022parameter}%
  \BibitemOpen
  \bibfield  {author} {\bibinfo {author} {\bibfnamefont {N.}~\bibnamefont {Christensen}}\ and\ \bibinfo {author} {\bibfnamefont {R.}~\bibnamefont {Meyer}},\ }\bibfield  {title} {\bibinfo {title} {Parameter estimation with gravitational waves},\ }\href@noop {} {\bibfield  {journal} {\bibinfo  {journal} {Reviews of Modern Physics}\ }\textbf {\bibinfo {volume} {94}},\ \bibinfo {pages} {025001} (\bibinfo {year} {2022})}\BibitemShut {NoStop}%
\bibitem [{\citenamefont {Jaranowski}\ and\ \citenamefont {Kr{\'o}lak}(2012)}]{jaranowski2012gravitational}%
  \BibitemOpen
  \bibfield  {author} {\bibinfo {author} {\bibfnamefont {P.}~\bibnamefont {Jaranowski}}\ and\ \bibinfo {author} {\bibfnamefont {A.}~\bibnamefont {Kr{\'o}lak}},\ }\bibfield  {title} {\bibinfo {title} {Gravitational-wave data analysis. formalism and sample applications: the gaussian case},\ }\href@noop {} {\bibfield  {journal} {\bibinfo  {journal} {Living Reviews in Relativity}\ }\textbf {\bibinfo {volume} {15}},\ \bibinfo {pages} {1} (\bibinfo {year} {2012})}\BibitemShut {NoStop}%
\bibitem [{\citenamefont {Smith}\ \emph {et~al.}(2021)\citenamefont {Smith}, \citenamefont {Borhanian}, \citenamefont {Sathyaprakash}, \citenamefont {Vivanco}, \citenamefont {Field}, \citenamefont {Lasky}, \citenamefont {Mandel}, \citenamefont {Morisaki}, \citenamefont {Ottaway}, \citenamefont {Slagmolen} \emph {et~al.}}]{smith2021bayesian}%
  \BibitemOpen
  \bibfield  {author} {\bibinfo {author} {\bibfnamefont {R.}~\bibnamefont {Smith}}, \bibinfo {author} {\bibfnamefont {S.}~\bibnamefont {Borhanian}}, \bibinfo {author} {\bibfnamefont {B.}~\bibnamefont {Sathyaprakash}}, \bibinfo {author} {\bibfnamefont {F.~H.}\ \bibnamefont {Vivanco}}, \bibinfo {author} {\bibfnamefont {S.~E.}\ \bibnamefont {Field}}, \bibinfo {author} {\bibfnamefont {P.}~\bibnamefont {Lasky}}, \bibinfo {author} {\bibfnamefont {I.}~\bibnamefont {Mandel}}, \bibinfo {author} {\bibfnamefont {S.}~\bibnamefont {Morisaki}}, \bibinfo {author} {\bibfnamefont {D.}~\bibnamefont {Ottaway}}, \bibinfo {author} {\bibfnamefont {B.~J.}\ \bibnamefont {Slagmolen}}, \emph {et~al.},\ }\bibfield  {title} {\bibinfo {title} {Bayesian inference for gravitational waves from binary neutron star mergers in third generation observatories},\ }\href@noop {} {\bibfield  {journal} {\bibinfo  {journal} {Physical review letters}\ }\textbf {\bibinfo {volume} {127}},\ \bibinfo {pages} {081102} (\bibinfo {year} {2021})}\BibitemShut
  {NoStop}%
\bibitem [{\citenamefont {Breschi}\ \emph {et~al.}(2021)\citenamefont {Breschi}, \citenamefont {Gamba},\ and\ \citenamefont {Bernuzzi}}]{breschi2021bayesian}%
  \BibitemOpen
  \bibfield  {author} {\bibinfo {author} {\bibfnamefont {M.}~\bibnamefont {Breschi}}, \bibinfo {author} {\bibfnamefont {R.}~\bibnamefont {Gamba}},\ and\ \bibinfo {author} {\bibfnamefont {S.}~\bibnamefont {Bernuzzi}},\ }\bibfield  {title} {\bibinfo {title} {Bayesian inference of multimessenger astrophysical data: Methods and applications to gravitational waves},\ }\href@noop {} {\bibfield  {journal} {\bibinfo  {journal} {Physical Review D}\ }\textbf {\bibinfo {volume} {104}},\ \bibinfo {pages} {042001} (\bibinfo {year} {2021})}\BibitemShut {NoStop}%
\bibitem [{\citenamefont {Chua}\ and\ \citenamefont {Vallisneri}(2020)}]{chua2020learning}%
  \BibitemOpen
  \bibfield  {author} {\bibinfo {author} {\bibfnamefont {A.~J.}\ \bibnamefont {Chua}}\ and\ \bibinfo {author} {\bibfnamefont {M.}~\bibnamefont {Vallisneri}},\ }\bibfield  {title} {\bibinfo {title} {Learning bayesian posteriors with neural networks for gravitational-wave inference},\ }\href@noop {} {\bibfield  {journal} {\bibinfo  {journal} {Physical review letters}\ }\textbf {\bibinfo {volume} {124}},\ \bibinfo {pages} {041102} (\bibinfo {year} {2020})}\BibitemShut {NoStop}%
\bibitem [{\citenamefont {Meyer}\ \emph {et~al.}(2022)\citenamefont {Meyer}, \citenamefont {Edwards}, \citenamefont {Maturana-Russel},\ and\ \citenamefont {Christensen}}]{meyer2022computational}%
  \BibitemOpen
  \bibfield  {author} {\bibinfo {author} {\bibfnamefont {R.}~\bibnamefont {Meyer}}, \bibinfo {author} {\bibfnamefont {M.~C.}\ \bibnamefont {Edwards}}, \bibinfo {author} {\bibfnamefont {P.}~\bibnamefont {Maturana-Russel}},\ and\ \bibinfo {author} {\bibfnamefont {N.}~\bibnamefont {Christensen}},\ }\bibfield  {title} {\bibinfo {title} {Computational techniques for parameter estimation of gravitational wave signals},\ }\href@noop {} {\bibfield  {journal} {\bibinfo  {journal} {Wiley Interdisciplinary Reviews: Computational Statistics}\ }\textbf {\bibinfo {volume} {14}},\ \bibinfo {pages} {e1532} (\bibinfo {year} {2022})}\BibitemShut {NoStop}%
\bibitem [{\citenamefont {Edwards}\ \emph {et~al.}(2014)\citenamefont {Edwards}, \citenamefont {Meyer},\ and\ \citenamefont {Christensen}}]{edwards2014bayesian}%
  \BibitemOpen
  \bibfield  {author} {\bibinfo {author} {\bibfnamefont {M.~C.}\ \bibnamefont {Edwards}}, \bibinfo {author} {\bibfnamefont {R.}~\bibnamefont {Meyer}},\ and\ \bibinfo {author} {\bibfnamefont {N.}~\bibnamefont {Christensen}},\ }\bibfield  {title} {\bibinfo {title} {Bayesian parameter estimation of core collapse supernovae using gravitational wave simulations},\ }\href@noop {} {\bibfield  {journal} {\bibinfo  {journal} {Inverse Problems}\ }\textbf {\bibinfo {volume} {30}},\ \bibinfo {pages} {114008} (\bibinfo {year} {2014})}\BibitemShut {NoStop}%
\bibitem [{\citenamefont {Dupuis}\ and\ \citenamefont {Woan}(2005)}]{dupuis2005bayesian}%
  \BibitemOpen
  \bibfield  {author} {\bibinfo {author} {\bibfnamefont {R.~J.}\ \bibnamefont {Dupuis}}\ and\ \bibinfo {author} {\bibfnamefont {G.}~\bibnamefont {Woan}},\ }\bibfield  {title} {\bibinfo {title} {Bayesian estimation of pulsar parameters from gravitational wave data},\ }\href@noop {} {\bibfield  {journal} {\bibinfo  {journal} {Physical Review D}\ }\textbf {\bibinfo {volume} {72}},\ \bibinfo {pages} {102002} (\bibinfo {year} {2005})}\BibitemShut {NoStop}%
\bibitem [{\citenamefont {Talbot}\ \emph {et~al.}(2019)\citenamefont {Talbot}, \citenamefont {Smith}, \citenamefont {Thrane},\ and\ \citenamefont {Poole}}]{talbot2019parallelized}%
  \BibitemOpen
  \bibfield  {author} {\bibinfo {author} {\bibfnamefont {C.}~\bibnamefont {Talbot}}, \bibinfo {author} {\bibfnamefont {R.}~\bibnamefont {Smith}}, \bibinfo {author} {\bibfnamefont {E.}~\bibnamefont {Thrane}},\ and\ \bibinfo {author} {\bibfnamefont {G.~B.}\ \bibnamefont {Poole}},\ }\bibfield  {title} {\bibinfo {title} {Parallelized inference for gravitational-wave astronomy},\ }\href@noop {} {\bibfield  {journal} {\bibinfo  {journal} {Physical Review D}\ }\textbf {\bibinfo {volume} {100}},\ \bibinfo {pages} {043030} (\bibinfo {year} {2019})}\BibitemShut {NoStop}%
\bibitem [{\citenamefont {Veitch}\ \emph {et~al.}(2015)\citenamefont {Veitch}, \citenamefont {Raymond}, \citenamefont {Farr}, \citenamefont {Farr}, \citenamefont {Graff}, \citenamefont {Vitale}, \citenamefont {Aylott}, \citenamefont {Blackburn}, \citenamefont {Christensen}, \citenamefont {Coughlin} \emph {et~al.}}]{veitch2015parameter}%
  \BibitemOpen
  \bibfield  {author} {\bibinfo {author} {\bibfnamefont {J.}~\bibnamefont {Veitch}}, \bibinfo {author} {\bibfnamefont {V.}~\bibnamefont {Raymond}}, \bibinfo {author} {\bibfnamefont {B.}~\bibnamefont {Farr}}, \bibinfo {author} {\bibfnamefont {W.}~\bibnamefont {Farr}}, \bibinfo {author} {\bibfnamefont {P.}~\bibnamefont {Graff}}, \bibinfo {author} {\bibfnamefont {S.}~\bibnamefont {Vitale}}, \bibinfo {author} {\bibfnamefont {B.}~\bibnamefont {Aylott}}, \bibinfo {author} {\bibfnamefont {K.}~\bibnamefont {Blackburn}}, \bibinfo {author} {\bibfnamefont {N.}~\bibnamefont {Christensen}}, \bibinfo {author} {\bibfnamefont {M.}~\bibnamefont {Coughlin}}, \emph {et~al.},\ }\bibfield  {title} {\bibinfo {title} {Parameter estimation for compact binaries with ground-based gravitational-wave observations using the lalinference software library},\ }\href@noop {} {\bibfield  {journal} {\bibinfo  {journal} {Physical Review D}\ }\textbf {\bibinfo {volume} {91}},\ \bibinfo {pages} {042003} (\bibinfo {year} {2015})}\BibitemShut
  {NoStop}%
\bibitem [{\citenamefont {Biwer}\ \emph {et~al.}(2019)\citenamefont {Biwer}, \citenamefont {Capano}, \citenamefont {De}, \citenamefont {Cabero}, \citenamefont {Brown}, \citenamefont {Nitz},\ and\ \citenamefont {Raymond}}]{biwer2019pycbc}%
  \BibitemOpen
  \bibfield  {author} {\bibinfo {author} {\bibfnamefont {C.~M.}\ \bibnamefont {Biwer}}, \bibinfo {author} {\bibfnamefont {C.~D.}\ \bibnamefont {Capano}}, \bibinfo {author} {\bibfnamefont {S.}~\bibnamefont {De}}, \bibinfo {author} {\bibfnamefont {M.}~\bibnamefont {Cabero}}, \bibinfo {author} {\bibfnamefont {D.~A.}\ \bibnamefont {Brown}}, \bibinfo {author} {\bibfnamefont {A.~H.}\ \bibnamefont {Nitz}},\ and\ \bibinfo {author} {\bibfnamefont {V.}~\bibnamefont {Raymond}},\ }\bibfield  {title} {\bibinfo {title} {Pycbc inference: A python-based parameter estimation toolkit for compact binary coalescence signals},\ }\href@noop {} {\bibfield  {journal} {\bibinfo  {journal} {Publications of the Astronomical Society of the Pacific}\ }\textbf {\bibinfo {volume} {131}},\ \bibinfo {pages} {024503} (\bibinfo {year} {2019})}\BibitemShut {NoStop}%
\bibitem [{\citenamefont {Ashton}\ \emph {et~al.}(2019)\citenamefont {Ashton}, \citenamefont {H{\"u}bner}, \citenamefont {Lasky}, \citenamefont {Talbot}, \citenamefont {Ackley}, \citenamefont {Biscoveanu}, \citenamefont {Chu}, \citenamefont {Divakarla}, \citenamefont {Easter}, \citenamefont {Goncharov} \emph {et~al.}}]{ashton2019bilby}%
  \BibitemOpen
  \bibfield  {author} {\bibinfo {author} {\bibfnamefont {G.}~\bibnamefont {Ashton}}, \bibinfo {author} {\bibfnamefont {M.}~\bibnamefont {H{\"u}bner}}, \bibinfo {author} {\bibfnamefont {P.~D.}\ \bibnamefont {Lasky}}, \bibinfo {author} {\bibfnamefont {C.}~\bibnamefont {Talbot}}, \bibinfo {author} {\bibfnamefont {K.}~\bibnamefont {Ackley}}, \bibinfo {author} {\bibfnamefont {S.}~\bibnamefont {Biscoveanu}}, \bibinfo {author} {\bibfnamefont {Q.}~\bibnamefont {Chu}}, \bibinfo {author} {\bibfnamefont {A.}~\bibnamefont {Divakarla}}, \bibinfo {author} {\bibfnamefont {P.~J.}\ \bibnamefont {Easter}}, \bibinfo {author} {\bibfnamefont {B.}~\bibnamefont {Goncharov}}, \emph {et~al.},\ }\bibfield  {title} {\bibinfo {title} {Bilby: A user-friendly bayesian inference library for gravitational-wave astronomy},\ }\href@noop {} {\bibfield  {journal} {\bibinfo  {journal} {The Astrophysical Journal Supplement Series}\ }\textbf {\bibinfo {volume} {241}},\ \bibinfo {pages} {27} (\bibinfo {year} {2019})}\BibitemShut {NoStop}%
\bibitem [{\citenamefont {Romero-Shaw}\ \emph {et~al.}(2020)\citenamefont {Romero-Shaw}, \citenamefont {Talbot}, \citenamefont {Biscoveanu}, \citenamefont {D’emilio}, \citenamefont {Ashton}, \citenamefont {Berry}, \citenamefont {Coughlin}, \citenamefont {Galaudage}, \citenamefont {Hoy}, \citenamefont {H{\"u}bner} \emph {et~al.}}]{romero2020bayesian}%
  \BibitemOpen
  \bibfield  {author} {\bibinfo {author} {\bibfnamefont {I.~M.}\ \bibnamefont {Romero-Shaw}}, \bibinfo {author} {\bibfnamefont {C.}~\bibnamefont {Talbot}}, \bibinfo {author} {\bibfnamefont {S.}~\bibnamefont {Biscoveanu}}, \bibinfo {author} {\bibfnamefont {V.}~\bibnamefont {D’emilio}}, \bibinfo {author} {\bibfnamefont {G.}~\bibnamefont {Ashton}}, \bibinfo {author} {\bibfnamefont {C.}~\bibnamefont {Berry}}, \bibinfo {author} {\bibfnamefont {S.}~\bibnamefont {Coughlin}}, \bibinfo {author} {\bibfnamefont {S.}~\bibnamefont {Galaudage}}, \bibinfo {author} {\bibfnamefont {C.}~\bibnamefont {Hoy}}, \bibinfo {author} {\bibfnamefont {M.}~\bibnamefont {H{\"u}bner}}, \emph {et~al.},\ }\bibfield  {title} {\bibinfo {title} {Bayesian inference for compact binary coalescences with bilby: validation and application to the first ligo--virgo gravitational-wave transient catalogue},\ }\href@noop {} {\bibfield  {journal} {\bibinfo  {journal} {Monthly Notices of the Royal Astronomical Society}\ }\textbf {\bibinfo {volume}
  {499}},\ \bibinfo {pages} {3295} (\bibinfo {year} {2020})}\BibitemShut {NoStop}%
\bibitem [{\citenamefont {Antil}\ \emph {et~al.}(2013)\citenamefont {Antil}, \citenamefont {Field}, \citenamefont {Herrmann}, \citenamefont {Nochetto},\ and\ \citenamefont {Tiglio}}]{antil2012two}%
  \BibitemOpen
  \bibfield  {author} {\bibinfo {author} {\bibfnamefont {H.}~\bibnamefont {Antil}}, \bibinfo {author} {\bibfnamefont {S.~E.}\ \bibnamefont {Field}}, \bibinfo {author} {\bibfnamefont {F.}~\bibnamefont {Herrmann}}, \bibinfo {author} {\bibfnamefont {R.~H.}\ \bibnamefont {Nochetto}},\ and\ \bibinfo {author} {\bibfnamefont {M.}~\bibnamefont {Tiglio}},\ }\bibfield  {title} {\bibinfo {title} {Two-step greedy algorithm for reduced order quadratures},\ }\href {https://doi.org/10.1007/s10915-013-9722-z} {\bibfield  {journal} {\bibinfo  {journal} {J. Sci. Comput.}\ }\textbf {\bibinfo {volume} {57}},\ \bibinfo {pages} {604} (\bibinfo {year} {2013})}\BibitemShut {NoStop}%
\bibitem [{\citenamefont {Canizares}\ \emph {et~al.}(2013)\citenamefont {Canizares}, \citenamefont {Field}, \citenamefont {Gair},\ and\ \citenamefont {Tiglio}}]{Canizares:2013ywa}%
  \BibitemOpen
  \bibfield  {author} {\bibinfo {author} {\bibfnamefont {P.}~\bibnamefont {Canizares}}, \bibinfo {author} {\bibfnamefont {S.~E.}\ \bibnamefont {Field}}, \bibinfo {author} {\bibfnamefont {J.~R.}\ \bibnamefont {Gair}},\ and\ \bibinfo {author} {\bibfnamefont {M.}~\bibnamefont {Tiglio}},\ }\bibfield  {title} {\bibinfo {title} {{Gravitational wave parameter estimation with compressed likelihood evaluations}},\ }\href {https://doi.org/10.1103/PhysRevD.87.124005} {\bibfield  {journal} {\bibinfo  {journal} {Phys. Rev.}\ }\textbf {\bibinfo {volume} {D87}},\ \bibinfo {pages} {124005} (\bibinfo {year} {2013})}\BibitemShut {NoStop}%
\bibitem [{\citenamefont {Canizares}\ \emph {et~al.}(2015)\citenamefont {Canizares}, \citenamefont {Field}, \citenamefont {Gair}, \citenamefont {Raymond}, \citenamefont {Smith},\ and\ \citenamefont {Tiglio}}]{canizares2015accelerated}%
  \BibitemOpen
  \bibfield  {author} {\bibinfo {author} {\bibfnamefont {P.}~\bibnamefont {Canizares}}, \bibinfo {author} {\bibfnamefont {S.~E.}\ \bibnamefont {Field}}, \bibinfo {author} {\bibfnamefont {J.}~\bibnamefont {Gair}}, \bibinfo {author} {\bibfnamefont {V.}~\bibnamefont {Raymond}}, \bibinfo {author} {\bibfnamefont {R.}~\bibnamefont {Smith}},\ and\ \bibinfo {author} {\bibfnamefont {M.}~\bibnamefont {Tiglio}},\ }\bibfield  {title} {\bibinfo {title} {Accelerated gravitational wave parameter estimation with reduced order modeling},\ }\href@noop {} {\bibfield  {journal} {\bibinfo  {journal} {Physical review letters}\ }\textbf {\bibinfo {volume} {114}},\ \bibinfo {pages} {071104} (\bibinfo {year} {2015})}\BibitemShut {NoStop}%
\bibitem [{\citenamefont {Barrault}\ \emph {et~al.}(2004)\citenamefont {Barrault}, \citenamefont {Maday}, \citenamefont {Nguyen},\ and\ \citenamefont {Patera}}]{barrault2004empirical}%
  \BibitemOpen
  \bibfield  {author} {\bibinfo {author} {\bibfnamefont {M.}~\bibnamefont {Barrault}}, \bibinfo {author} {\bibfnamefont {Y.}~\bibnamefont {Maday}}, \bibinfo {author} {\bibfnamefont {N.~C.}\ \bibnamefont {Nguyen}},\ and\ \bibinfo {author} {\bibfnamefont {A.~T.}\ \bibnamefont {Patera}},\ }\bibfield  {title} {\bibinfo {title} {An ‘empirical interpolation’method: application to efficient reduced-basis discretization of partial differential equations},\ }\href@noop {} {\bibfield  {journal} {\bibinfo  {journal} {Comptes Rendus Mathematique}\ }\textbf {\bibinfo {volume} {339}},\ \bibinfo {pages} {667} (\bibinfo {year} {2004})}\BibitemShut {NoStop}%
\bibitem [{\citenamefont {Dax}\ \emph {et~al.}(2021)\citenamefont {Dax}, \citenamefont {Green}, \citenamefont {Gair}, \citenamefont {Macke}, \citenamefont {Buonanno},\ and\ \citenamefont {Sch{\"o}lkopf}}]{dax2021real}%
  \BibitemOpen
  \bibfield  {author} {\bibinfo {author} {\bibfnamefont {M.}~\bibnamefont {Dax}}, \bibinfo {author} {\bibfnamefont {S.~R.}\ \bibnamefont {Green}}, \bibinfo {author} {\bibfnamefont {J.}~\bibnamefont {Gair}}, \bibinfo {author} {\bibfnamefont {J.~H.}\ \bibnamefont {Macke}}, \bibinfo {author} {\bibfnamefont {A.}~\bibnamefont {Buonanno}},\ and\ \bibinfo {author} {\bibfnamefont {B.}~\bibnamefont {Sch{\"o}lkopf}},\ }\bibfield  {title} {\bibinfo {title} {Real-time gravitational wave science with neural posterior estimation},\ }\href@noop {} {\bibfield  {journal} {\bibinfo  {journal} {Physical review letters}\ }\textbf {\bibinfo {volume} {127}},\ \bibinfo {pages} {241103} (\bibinfo {year} {2021})}\BibitemShut {NoStop}%
\bibitem [{\citenamefont {Gabbard}\ \emph {et~al.}(2019)\citenamefont {Gabbard}, \citenamefont {Messenger}, \citenamefont {Heng}, \citenamefont {Tonolini},\ and\ \citenamefont {Murray-Smith}}]{gabbard2019bayesian}%
  \BibitemOpen
  \bibfield  {author} {\bibinfo {author} {\bibfnamefont {H.}~\bibnamefont {Gabbard}}, \bibinfo {author} {\bibfnamefont {C.}~\bibnamefont {Messenger}}, \bibinfo {author} {\bibfnamefont {I.~S.}\ \bibnamefont {Heng}}, \bibinfo {author} {\bibfnamefont {F.}~\bibnamefont {Tonolini}},\ and\ \bibinfo {author} {\bibfnamefont {R.}~\bibnamefont {Murray-Smith}},\ }\bibfield  {title} {\bibinfo {title} {Bayesian parameter estimation using conditional variational autoencoders for gravitational-wave astronomy. arxiv e-prints},\ }\href@noop {} {\bibfield  {journal} {\bibinfo  {journal} {arXiv preprint arXiv:1909.06296}\ } (\bibinfo {year} {2019})}\BibitemShut {NoStop}%
\bibitem [{\citenamefont {Green}\ and\ \citenamefont {Gair}(2020)}]{green2020complete}%
  \BibitemOpen
  \bibfield  {author} {\bibinfo {author} {\bibfnamefont {S.}~\bibnamefont {Green}}\ and\ \bibinfo {author} {\bibfnamefont {J.}~\bibnamefont {Gair}},\ }\bibfield  {title} {\bibinfo {title} {Complete parameter inference for gw150914 using deep learning. arxiv e-prints},\ }\href@noop {} {\bibfield  {journal} {\bibinfo  {journal} {arXiv preprint arXiv:2008.03312}\ } (\bibinfo {year} {2020})}\BibitemShut {NoStop}%
\bibitem [{\citenamefont {Green}\ \emph {et~al.}(2020)\citenamefont {Green}, \citenamefont {Simpson},\ and\ \citenamefont {Gair}}]{green2020gravitational}%
  \BibitemOpen
  \bibfield  {author} {\bibinfo {author} {\bibfnamefont {S.~R.}\ \bibnamefont {Green}}, \bibinfo {author} {\bibfnamefont {C.}~\bibnamefont {Simpson}},\ and\ \bibinfo {author} {\bibfnamefont {J.}~\bibnamefont {Gair}},\ }\bibfield  {title} {\bibinfo {title} {Gravitational-wave parameter estimation with autoregressive neural network flows},\ }\href@noop {} {\bibfield  {journal} {\bibinfo  {journal} {Physical Review D}\ }\textbf {\bibinfo {volume} {102}},\ \bibinfo {pages} {104057} (\bibinfo {year} {2020})}\BibitemShut {NoStop}%
\bibitem [{\citenamefont {George}\ and\ \citenamefont {Huerta}(2017)}]{george2017deep}%
  \BibitemOpen
  \bibfield  {author} {\bibinfo {author} {\bibfnamefont {D.}~\bibnamefont {George}}\ and\ \bibinfo {author} {\bibfnamefont {E.}~\bibnamefont {Huerta}},\ }\bibfield  {title} {\bibinfo {title} {Deep learning for real-time gravitational wave detection and parameter estimation with ligo data},\ }\href@noop {} {\bibfield  {journal} {\bibinfo  {journal} {arXiv preprint arXiv:1711.07966}\ } (\bibinfo {year} {2017})}\BibitemShut {NoStop}%
\bibitem [{\citenamefont {{\'A}lvares}\ \emph {et~al.}(2021)\citenamefont {{\'A}lvares}, \citenamefont {Font}, \citenamefont {Freitas}, \citenamefont {Freitas}, \citenamefont {Morais}, \citenamefont {Nunes}, \citenamefont {Onofre},\ and\ \citenamefont {Torres-Forn{\'e}}}]{alvares2021exploring}%
  \BibitemOpen
  \bibfield  {author} {\bibinfo {author} {\bibfnamefont {J.~D.}\ \bibnamefont {{\'A}lvares}}, \bibinfo {author} {\bibfnamefont {J.~A.}\ \bibnamefont {Font}}, \bibinfo {author} {\bibfnamefont {F.~F.}\ \bibnamefont {Freitas}}, \bibinfo {author} {\bibfnamefont {O.~G.}\ \bibnamefont {Freitas}}, \bibinfo {author} {\bibfnamefont {A.~P.}\ \bibnamefont {Morais}}, \bibinfo {author} {\bibfnamefont {S.}~\bibnamefont {Nunes}}, \bibinfo {author} {\bibfnamefont {A.}~\bibnamefont {Onofre}},\ and\ \bibinfo {author} {\bibfnamefont {A.}~\bibnamefont {Torres-Forn{\'e}}},\ }\bibfield  {title} {\bibinfo {title} {Exploring gravitational-wave detection and parameter inference using deep learning methods},\ }\href@noop {} {\bibfield  {journal} {\bibinfo  {journal} {Classical and Quantum Gravity}\ }\textbf {\bibinfo {volume} {38}},\ \bibinfo {pages} {155010} (\bibinfo {year} {2021})}\BibitemShut {NoStop}%
\bibitem [{\citenamefont {Shen}\ \emph {et~al.}(2021)\citenamefont {Shen}, \citenamefont {Huerta}, \citenamefont {O’Shea}, \citenamefont {Kumar},\ and\ \citenamefont {Zhao}}]{shen2021statistically}%
  \BibitemOpen
  \bibfield  {author} {\bibinfo {author} {\bibfnamefont {H.}~\bibnamefont {Shen}}, \bibinfo {author} {\bibfnamefont {E.}~\bibnamefont {Huerta}}, \bibinfo {author} {\bibfnamefont {E.}~\bibnamefont {O’Shea}}, \bibinfo {author} {\bibfnamefont {P.}~\bibnamefont {Kumar}},\ and\ \bibinfo {author} {\bibfnamefont {Z.}~\bibnamefont {Zhao}},\ }\bibfield  {title} {\bibinfo {title} {Statistically-informed deep learning for gravitational wave parameter estimation},\ }\href@noop {} {\bibfield  {journal} {\bibinfo  {journal} {Machine Learning: Science and Technology}\ }\textbf {\bibinfo {volume} {3}},\ \bibinfo {pages} {015007} (\bibinfo {year} {2021})}\BibitemShut {NoStop}%
\bibitem [{\citenamefont {Morisaki}\ and\ \citenamefont {Raymond}(2020)}]{Morisaki_2020}%
  \BibitemOpen
  \bibfield  {author} {\bibinfo {author} {\bibfnamefont {S.}~\bibnamefont {Morisaki}}\ and\ \bibinfo {author} {\bibfnamefont {V.}~\bibnamefont {Raymond}},\ }\bibfield  {title} {\bibinfo {title} {{Rapid Parameter Estimation of Gravitational Waves from Binary Neutron Star Coalescence using Focused Reduced Order Quadrature}},\ }\href {https://doi.org/10.1103/PhysRevD.102.104020} {\bibfield  {journal} {\bibinfo  {journal} {Phys. Rev. D}\ }\textbf {\bibinfo {volume} {102}},\ \bibinfo {pages} {104020} (\bibinfo {year} {2020})}\BibitemShut {NoStop}%
\bibitem [{\citenamefont {Eftang}(2011)}]{EftangThesis}%
  \BibitemOpen
  \bibfield  {author} {\bibinfo {author} {\bibfnamefont {J.~L.}\ \bibnamefont {Eftang}},\ }\bibfield  {title} {\bibinfo {title} {Reduced basis methods for parametrized partial differential equations},\ }\href@noop {} {\bibfield  {journal} {\bibinfo  {journal} {Norwegian University of Science and Technology}\ } (\bibinfo {year} {2011})}\BibitemShut {NoStop}%
\bibitem [{\citenamefont {Caudill}\ \emph {et~al.}(2012)\citenamefont {Caudill}, \citenamefont {Field}, \citenamefont {Galley}, \citenamefont {Herrmann},\ and\ \citenamefont {Tiglio}}]{Caudill:2011kv}%
  \BibitemOpen
  \bibfield  {author} {\bibinfo {author} {\bibfnamefont {S.}~\bibnamefont {Caudill}}, \bibinfo {author} {\bibfnamefont {S.~E.}\ \bibnamefont {Field}}, \bibinfo {author} {\bibfnamefont {C.~R.}\ \bibnamefont {Galley}}, \bibinfo {author} {\bibfnamefont {F.}~\bibnamefont {Herrmann}},\ and\ \bibinfo {author} {\bibfnamefont {M.}~\bibnamefont {Tiglio}},\ }\bibfield  {title} {\bibinfo {title} {{Reduced Basis representations of multi-mode black hole ringdown gravitational waves}},\ }\href {https://iopscience.iop.org/article/10.1088/0264-9381/29/9/095016} {\bibfield  {journal} {\bibinfo  {journal} {Class. Quant. Grav.}\ }\textbf {\bibinfo {volume} {29}},\ \bibinfo {pages} {095016} (\bibinfo {year} {2012})}\BibitemShut {NoStop}%
\bibitem [{\citenamefont {Bergstra}\ \emph {et~al.}(2011)\citenamefont {Bergstra}, \citenamefont {Bardenet}, \citenamefont {Bengio},\ and\ \citenamefont {K\'{e}gl}}]{NIPS2011_86e8f7ab}%
  \BibitemOpen
  \bibfield  {author} {\bibinfo {author} {\bibfnamefont {J.}~\bibnamefont {Bergstra}}, \bibinfo {author} {\bibfnamefont {R.}~\bibnamefont {Bardenet}}, \bibinfo {author} {\bibfnamefont {Y.}~\bibnamefont {Bengio}},\ and\ \bibinfo {author} {\bibfnamefont {B.}~\bibnamefont {K\'{e}gl}},\ }\bibfield  {title} {\bibinfo {title} {Algorithms for hyper-parameter optimization},\ }in\ \href {https://proceedings.neurips.cc/paper/2011/file/86e8f7ab32cfd12577bc2619bc635690-Paper.pdf} {\emph {\bibinfo {booktitle} {Advances in Neural Information Processing Systems}}},\ Vol.~\bibinfo {volume} {24},\ \bibinfo {editor} {edited by\ \bibinfo {editor} {\bibfnamefont {J.}~\bibnamefont {Shawe-Taylor}}, \bibinfo {editor} {\bibfnamefont {R.}~\bibnamefont {Zemel}}, \bibinfo {editor} {\bibfnamefont {P.}~\bibnamefont {Bartlett}}, \bibinfo {editor} {\bibfnamefont {F.}~\bibnamefont {Pereira}},\ and\ \bibinfo {editor} {\bibfnamefont {K.}~\bibnamefont {Weinberger}}}\ (\bibinfo  {publisher} {Curran Associates, Inc.},\ \bibinfo {year}
  {2011})\BibitemShut {NoStop}%
\bibitem [{\citenamefont {Akiba}\ \emph {et~al.}(2019)\citenamefont {Akiba}, \citenamefont {Sano}, \citenamefont {Yanase}, \citenamefont {Ohta},\ and\ \citenamefont {Koyama}}]{optuna_2019}%
  \BibitemOpen
  \bibfield  {author} {\bibinfo {author} {\bibfnamefont {T.}~\bibnamefont {Akiba}}, \bibinfo {author} {\bibfnamefont {S.}~\bibnamefont {Sano}}, \bibinfo {author} {\bibfnamefont {T.}~\bibnamefont {Yanase}}, \bibinfo {author} {\bibfnamefont {T.}~\bibnamefont {Ohta}},\ and\ \bibinfo {author} {\bibfnamefont {M.}~\bibnamefont {Koyama}},\ }\bibfield  {title} {\bibinfo {title} {Optuna: A next-generation hyperparameter optimization framework},\ }in\ \href@noop {} {\emph {\bibinfo {booktitle} {Proceedings of the 25rd {ACM} {SIGKDD} International Conference on Knowledge Discovery and Data Mining}}}\ (\bibinfo {year} {2019})\BibitemShut {NoStop}%
\bibitem [{\citenamefont {Binev}\ \emph {et~al.}(2011)\citenamefont {Binev}, \citenamefont {Cohen}, \citenamefont {Dahmen}, \citenamefont {DeVore}, \citenamefont {Petrova},\ and\ \citenamefont {Wojtaszczyk}}]{Binev10convergencerates}%
  \BibitemOpen
  \bibfield  {author} {\bibinfo {author} {\bibfnamefont {P.}~\bibnamefont {Binev}}, \bibinfo {author} {\bibfnamefont {A.}~\bibnamefont {Cohen}}, \bibinfo {author} {\bibfnamefont {W.}~\bibnamefont {Dahmen}}, \bibinfo {author} {\bibfnamefont {R.~A.}\ \bibnamefont {DeVore}}, \bibinfo {author} {\bibfnamefont {G.}~\bibnamefont {Petrova}},\ and\ \bibinfo {author} {\bibfnamefont {P.}~\bibnamefont {Wojtaszczyk}},\ }\bibfield  {title} {\bibinfo {title} {Convergence rates for greedy algorithms in reduced basis methods},\ }\href {https://www.researchgate.net/publication/220132479_Convergence_Rates_for_Greedy_Algorithms_in_Reduced_Basis_Methods} {\bibfield  {journal} {\bibinfo  {journal} {SIAM J. Math. Analysis}\ }\textbf {\bibinfo {volume} {43}},\ \bibinfo {pages} {1457} (\bibinfo {year} {2011})}\BibitemShut {NoStop}%
\bibitem [{\citenamefont {DeVore}\ \emph {et~al.}(2013)\citenamefont {DeVore}, \citenamefont {Petrova},\ and\ \citenamefont {Wojtaszczyk}}]{Devore2012}%
  \BibitemOpen
  \bibfield  {author} {\bibinfo {author} {\bibfnamefont {R.}~\bibnamefont {DeVore}}, \bibinfo {author} {\bibfnamefont {G.}~\bibnamefont {Petrova}},\ and\ \bibinfo {author} {\bibfnamefont {P.}~\bibnamefont {Wojtaszczyk}},\ }\bibfield  {title} {\bibinfo {title} {Greedy algorithms for reduced bases in banach spaces},\ }\href {https://doi.org/10.1007/s00365-013-9186-2} {\bibfield  {journal} {\bibinfo  {journal} {Constructive Approximation}\ }\textbf {\bibinfo {volume} {37}},\ \bibinfo {pages} {455} (\bibinfo {year} {2013})}\BibitemShut {NoStop}%
\bibitem [{\citenamefont {Karniadakis}\ and\ \citenamefont {Sherwin}(2005)}]{karniadakis2005spectralhp}%
  \BibitemOpen
  \bibfield  {author} {\bibinfo {author} {\bibfnamefont {G.}~\bibnamefont {Karniadakis}}\ and\ \bibinfo {author} {\bibfnamefont {S.~J.}\ \bibnamefont {Sherwin}},\ }\href {https://doi.org/10.1093/acprof:oso/9780198528692.001.0001} {\emph {\bibinfo {title} {Spectral/hp Element Methods for Computational Fluid Dynamics}}},\ \bibinfo {edition} {2nd}\ ed.\ (\bibinfo  {publisher} {Oxford University Press},\ \bibinfo {address} {Oxford},\ \bibinfo {year} {2005})\BibitemShut {NoStop}%
\bibitem [{\citenamefont {Shahriari}\ \emph {et~al.}(2016)\citenamefont {Shahriari}, \citenamefont {Swersky}, \citenamefont {Wang}, \citenamefont {Adams},\ and\ \citenamefont {de~Freitas}}]{7352306}%
  \BibitemOpen
  \bibfield  {author} {\bibinfo {author} {\bibfnamefont {B.}~\bibnamefont {Shahriari}}, \bibinfo {author} {\bibfnamefont {K.}~\bibnamefont {Swersky}}, \bibinfo {author} {\bibfnamefont {Z.}~\bibnamefont {Wang}}, \bibinfo {author} {\bibfnamefont {R.~P.}\ \bibnamefont {Adams}},\ and\ \bibinfo {author} {\bibfnamefont {N.}~\bibnamefont {de~Freitas}},\ }\bibfield  {title} {\bibinfo {title} {Taking the human out of the loop: A review of bayesian optimization},\ }\href {https://doi.org/10.1109/JPROC.2015.2494218} {\bibfield  {journal} {\bibinfo  {journal} {Proceedings of the IEEE}\ }\textbf {\bibinfo {volume} {104}},\ \bibinfo {pages} {148} (\bibinfo {year} {2016})}\BibitemShut {NoStop}%
\bibitem [{\citenamefont {Brochu}\ \emph {et~al.}(2010)\citenamefont {Brochu}, \citenamefont {Cora},\ and\ \citenamefont {de~Freitas}}]{https://doi.org/10.48550/arxiv.1012.2599}%
  \BibitemOpen
  \bibfield  {author} {\bibinfo {author} {\bibfnamefont {E.}~\bibnamefont {Brochu}}, \bibinfo {author} {\bibfnamefont {V.~M.}\ \bibnamefont {Cora}},\ and\ \bibinfo {author} {\bibfnamefont {N.}~\bibnamefont {de~Freitas}},\ }\href {https://doi.org/10.48550/ARXIV.1012.2599} {\bibinfo {title} {A tutorial on bayesian optimization of expensive cost functions, with application to active user modeling and hierarchical reinforcement learning}} (\bibinfo {year} {2010})\BibitemShut {NoStop}%
\bibitem [{\citenamefont {Dewancker}\ \emph {et~al.}(2015)\citenamefont {Dewancker}, \citenamefont {McCourt},\ and\ \citenamefont {Clark}}]{dewancker2015bayesian}%
  \BibitemOpen
  \bibfield  {author} {\bibinfo {author} {\bibfnamefont {I.}~\bibnamefont {Dewancker}}, \bibinfo {author} {\bibfnamefont {M.}~\bibnamefont {McCourt}},\ and\ \bibinfo {author} {\bibfnamefont {S.}~\bibnamefont {Clark}},\ }\href {https://app.sigopt.com/static/pdf/SigOpt_Bayesian_Optimization_Primer.pdf} {\bibinfo {title} {Bayesian optimization primer}} (\bibinfo {year} {2015})\BibitemShut {NoStop}%
\bibitem [{\citenamefont {Feurer}\ and\ \citenamefont {Hutter}(2019)}]{Feurer2019}%
  \BibitemOpen
  \bibfield  {author} {\bibinfo {author} {\bibfnamefont {M.}~\bibnamefont {Feurer}}\ and\ \bibinfo {author} {\bibfnamefont {F.}~\bibnamefont {Hutter}},\ }\bibinfo {title} {Hyperparameter optimization},\ in\ \href {https://doi.org/10.1007/978-3-030-05318-5_1} {\emph {\bibinfo {booktitle} {Automated Machine Learning: Methods, Systems, Challenges}}},\ \bibinfo {editor} {edited by\ \bibinfo {editor} {\bibfnamefont {F.}~\bibnamefont {Hutter}}, \bibinfo {editor} {\bibfnamefont {L.}~\bibnamefont {Kotthoff}},\ and\ \bibinfo {editor} {\bibfnamefont {J.}~\bibnamefont {Vanschoren}}}\ (\bibinfo  {publisher} {Springer International Publishing},\ \bibinfo {address} {Cham},\ \bibinfo {year} {2019})\ pp.\ \bibinfo {pages} {3--33}\BibitemShut {NoStop}%
\bibitem [{\citenamefont {Jones}(2001)}]{EI1}%
  \BibitemOpen
  \bibfield  {author} {\bibinfo {author} {\bibfnamefont {D.}~\bibnamefont {Jones}},\ }\bibfield  {title} {\bibinfo {title} {A taxonomy of global optimization methods based on response surfaces},\ }\href {https://doi.org/10.1023/A:1012771025575} {\bibfield  {journal} {\bibinfo  {journal} {J. of Global Optimization}\ }\textbf {\bibinfo {volume} {21}},\ \bibinfo {pages} {345} (\bibinfo {year} {2001})}\BibitemShut {NoStop}%
\bibitem [{\citenamefont {Parzen}(1962)}]{parzen1962}%
  \BibitemOpen
  \bibfield  {author} {\bibinfo {author} {\bibfnamefont {E.}~\bibnamefont {Parzen}},\ }\bibfield  {title} {\bibinfo {title} {{On Estimation of a Probability Density Function and Mode}},\ }\href {https://doi.org/10.1214/aoms/1177704472} {\bibfield  {journal} {\bibinfo  {journal} {The Annals of Mathematical Statistics}\ }\textbf {\bibinfo {volume} {33}},\ \bibinfo {pages} {1065 } (\bibinfo {year} {1962})}\BibitemShut {NoStop}%
\bibitem [{\citenamefont {Ozaki}\ \emph {et~al.}(2022)\citenamefont {Ozaki}, \citenamefont {Tanigaki}, \citenamefont {Watanabe}, \citenamefont {Nomura},\ and\ \citenamefont {Onishi}}]{10.1613/jair.1.13188}%
  \BibitemOpen
  \bibfield  {author} {\bibinfo {author} {\bibfnamefont {Y.}~\bibnamefont {Ozaki}}, \bibinfo {author} {\bibfnamefont {Y.}~\bibnamefont {Tanigaki}}, \bibinfo {author} {\bibfnamefont {S.}~\bibnamefont {Watanabe}}, \bibinfo {author} {\bibfnamefont {M.}~\bibnamefont {Nomura}},\ and\ \bibinfo {author} {\bibfnamefont {M.}~\bibnamefont {Onishi}},\ }\bibfield  {title} {\bibinfo {title} {Multiobjective tree-structured parzen estimator},\ }\bibfield  {journal} {\bibinfo  {journal} {J. Artif. Int. Res.}\ }\textbf {\bibinfo {volume} {73}},\ \href {https://doi.org/10.1613/jair.1.13188} {10.1613/jair.1.13188} (\bibinfo {year} {2022})\BibitemShut {NoStop}%
\bibitem [{\citenamefont {Varma}\ \emph {et~al.}(2019)\citenamefont {Varma}, \citenamefont {Field}, \citenamefont {Scheel}, \citenamefont {Blackman}, \citenamefont {Kidder},\ and\ \citenamefont {Pfeiffer}}]{varma2019surrogate}%
  \BibitemOpen
  \bibfield  {author} {\bibinfo {author} {\bibfnamefont {V.}~\bibnamefont {Varma}}, \bibinfo {author} {\bibfnamefont {S.~E.}\ \bibnamefont {Field}}, \bibinfo {author} {\bibfnamefont {M.~A.}\ \bibnamefont {Scheel}}, \bibinfo {author} {\bibfnamefont {J.}~\bibnamefont {Blackman}}, \bibinfo {author} {\bibfnamefont {L.~E.}\ \bibnamefont {Kidder}},\ and\ \bibinfo {author} {\bibfnamefont {H.~P.}\ \bibnamefont {Pfeiffer}},\ }\bibfield  {title} {\bibinfo {title} {Surrogate model of hybridized numerical relativity binary black hole waveforms},\ }\href@noop {} {\bibfield  {journal} {\bibinfo  {journal} {Physical Review D}\ }\textbf {\bibinfo {volume} {99}},\ \bibinfo {pages} {064045} (\bibinfo {year} {2019})}\BibitemShut {NoStop}%
\bibitem [{\citenamefont {Villegas}(2023)}]{hp-greedy-bayesian-optimization}%
  \BibitemOpen
  \bibfield  {author} {\bibinfo {author} {\bibfnamefont {A.}~\bibnamefont {Villegas}},\ }\href@noop {} {\bibinfo {title} {hp-greedy bayesian optimization}},\ \bibinfo {howpublished} {\url{https://github.com/atuel96/hp-greedy-bayesian-optimization}} (\bibinfo {year} {2023})\BibitemShut {NoStop}%
\bibitem [{\citenamefont {Cerino}(2022)}]{scikit-reducedmodel}%
  \BibitemOpen
  \bibfield  {author} {\bibinfo {author} {\bibfnamefont {F.}~\bibnamefont {Cerino}},\ }\href@noop {} {\bibinfo {title} {Scikit-reducedmodel}},\ \bibinfo {howpublished} {\url{https://github.com/francocerino/scikit-reducedmodel}} (\bibinfo {year} {2022})\BibitemShut {NoStop}%
\end{thebibliography}%

\end{document}